\pdfoutput=1

\documentclass[11pt,twoside,a4paper,cmspaper,final,collab]{cms-tdr}
\def\svnVersion{260982}\def\cmsCernNo{2013-037}\def\cmsCernDate{\today}\def\cmsMessage{Published in the European Physical Journal C as \href{http://dx.doi.org/10.1140/epjc/s10052-014-3060-7}{\doi{10.1140/epjc/s10052-014-3060-7.}}}

\begin{document}\cmsNoteHeader{TOP-12-036}

\hyphenation{had-ron-i-za-tion}
\hyphenation{cal-or-i-me-ter}
\hyphenation{de-vices}
\RCS$Revision: 260114 $
\RCS$HeadURL: svn+ssh://svn.cern.ch/reps/tdr2/papers/TOP-12-036/trunk/TOP-12-036.tex $
\RCS$Id: TOP-12-036.tex 260114 2014-09-12 11:29:12Z mangano $
\newlength\cmsFigWidth
\ifthenelse{\boolean{cms@external}}{\setlength\cmsFigWidth{0.95\columnwidth}}{\setlength\cmsFigWidth{0.4\textwidth}}
\newlength\cmsFigWidthLarge
\ifthenelse{\boolean{cms@external}}{\setlength\cmsFigWidthLarge{0.95\columnwidth}}{\setlength\cmsFigWidthLarge{0.6\textwidth}}
\newlength\cmsFigWidthFeyn
\ifthenelse{\boolean{cms@external}}{\setlength\cmsFigWidthFeyn{0.7\columnwidth}}{\setlength\cmsFigWidthFeyn{0.4\textwidth}}
\ifthenelse{\boolean{cms@external}}{\providecommand{\cmsLeft}{top}}{\providecommand{\cmsLeft}{left}}
\ifthenelse{\boolean{cms@external}}{\providecommand{\cmsRight}{bottom}}{\providecommand{\cmsRight}{right}}
\newcommand{\V}{\ensuremath{\cmsSymbolFace{V}}\xspace}
\newcommand{\ttw}{\ensuremath{\ttbar\PW}\xspace}
\newcommand{\ttz}{\ensuremath{\ttbar\cPZ}\xspace}
\newcommand{\ttv}{\ensuremath{\ttbar\V}\xspace}
\newcommand{\ttg}{\ensuremath{\ttbar\gamma}\xspace}
\newcommand{\ttH}{\ensuremath{\ttbar\PH}\xspace}
\newcommand{\epem}{\Pep\Pem\xspace}
\newcommand{\Pepm}{\ensuremath{\Pe^\pm}}
\newcommand{\wjets}{\ensuremath{\PW\text{+jets}}\xspace}
\newcommand{\lumiValue}{19.5\fbinv}
\providecommand{\W}{\ensuremath{\PW}\xspace}
\providecommand{\systOstat}{\ensuremath{\,(\text{syst}\oplus\text{stat})}\xspace}
\titlerunning{Measurement of \ttbar with $\PW$ or $\cPZ$}
\cmsNoteHeader{TOP-12-036}
\title{Measurement of top quark-antiquark pair production in association with a W or Z boson in pp collisions at $\sqrt{s} = 8$\TeV}

\date{\today}

\abstract{
A measurement of the cross section for the production of top quark-antiquark pairs~(\ttbar) in association with a vector
boson V (W or Z) in
proton-proton collisions at $\sqrt{s} = 8$\TeV is presented. The results are based on a
dataset corresponding to an integrated luminosity of 19.5\fbinv recorded with the CMS detector at the LHC.
The measurement is performed in three leptonic (e and $\mu$) channels: a same-sign dilepton analysis targeting $\ttw$
events, and trilepton and four-lepton analyses designed for $\ttz$ events. In the same-sign dilepton channel, the $\ttw$ cross section
is measured as
$\sigma_{\ttw} = 170 ^{+90}_{-80}\stat \pm 70\syst\unit{fb}$,
corresponding to a significance of  1.6 standard deviations over the background-only hypothesis.
Combining the trilepton and four-lepton channels, a direct measurement of the $\ttz$ cross section,
$\sigma_{\ttz} = 200  ^{+80}_{-70}\stat ^{+40}_{-30}\syst\unit{fb}$,
is obtained with a significance of  3.1 standard deviations. The measured cross sections are compatible with standard model
predictions within their experimental uncertainties. The
inclusive $\ttv$ process is observed with a significance of 3.7 standard deviations from the combination of all three leptonic channels.
}

\hypersetup{%
pdfauthor={CMS Collaboration},%
pdftitle={Measurement of top quark-antiquark pair production in association with a W or Z boson in pp collisions at sqrt(s) = 8 TeV},%
pdfsubject={CMS},%
pdfkeywords={CMS, physics, top quark}}

\maketitle %maketitle comes after all the front information has been supplied

\section{Introduction}
Two decades after the discovery of the top quark~\cite{top_cdf,top_d0_discovery}, many of its properties
are still to be determined or are
only loosely constrained by experimental data. Among these properties are the couplings between
the top quark and the vector bosons.

The existence of non-zero
couplings between the top quark and the neutral vector bosons can be inferred
through the analysis of direct production of \ttbar pairs in association with a $\gamma$ or a \Z boson.
The CERN LHC allows these two processes to be disentangled and
the corresponding couplings to be measured.
The associated production of \ttbar pairs with
a W boson, the \ttw process,
has a cross section similar to \ttz and \ttg production. All three processes can be used
to test the internal consistency of the standard model (SM)~\cite{sm_Glashow,sm_Weinberg,sm_Salam} and search for the presence
of new physics.
Despite their small cross sections, they
are significant backgrounds to analyses that probe phenomena with even
smaller, or comparable, cross sections. Examples are
searches for supersymmetry~\cite{Barnett:1993ea,Guchait:1994zk,Baer:1995va} in same-sign
dilepton~\cite{ssDilepSusy8TeV} and in multilepton~\cite{multileptonSearch7TeV} final states, and
the analysis of the SM \ttH process with the Higgs boson and the top quarks decaying leptonically.

The measurement of the \ttg process has been documented by the CDF Collaboration~\cite{ttgCDF} for
proton-antiproton collisions at a centre-of-mass energy $\sqrt{s}=1.96\TeV$.
This article presents instead the measurement of cross sections for the \ttw and \ttz processes
in proton-proton (pp) collisions at $\sqrt{s}=8\TeV$.
The analysis is based on data corresponding to an integrated luminosity of \lumiValue
collected with the CMS detector at the LHC in 2012.
Unlike the previous observation of
the \ttv  process (\V equal to \W or \Z) at $\sqrt{s}=7\TeV$~\cite{ttVat7TeV}, here the \ttw process
is treated separately.

Three leptonic (\Pe~and \Pgm) final states are considered: same-sign dilepton events, trilepton events, and
four-lepton events. The same-sign dilepton events are used for the measurement
of the \ttw process, where one lepton originates from the leptonic decay of one of the two top
quarks and the other like-sign lepton is produced in the decay of the prompt vector boson. The trilepton
events are used for the identification of \ttz events in which one lepton is again produced from
the leptonic decay of one of the two top quarks, and the two other opposite-sign and
same-flavour leptons stem from the decay of the Z boson. The four-lepton events are used to
identify \ttz events in
which both the top quarks and the Z boson decay leptonically. For all three signatures,
signal events containing leptonic $\tau$ decays are implicitly included.

\begin{figure}[htb]
\centering
\includegraphics[width=\cmsFigWidthFeyn]{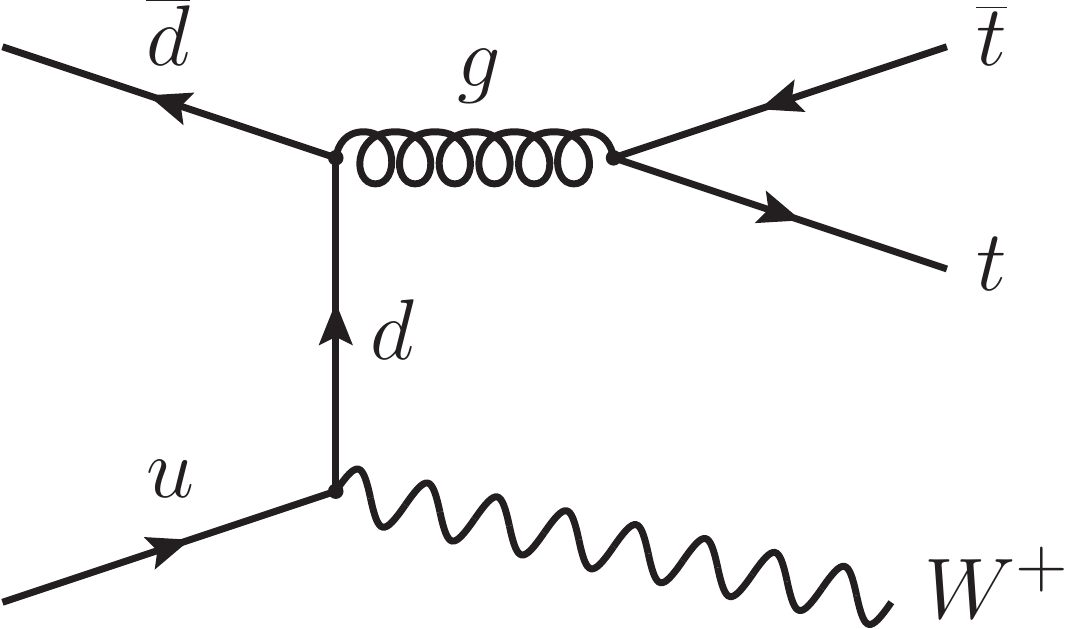}\hfil
\includegraphics[width=\cmsFigWidthFeyn]{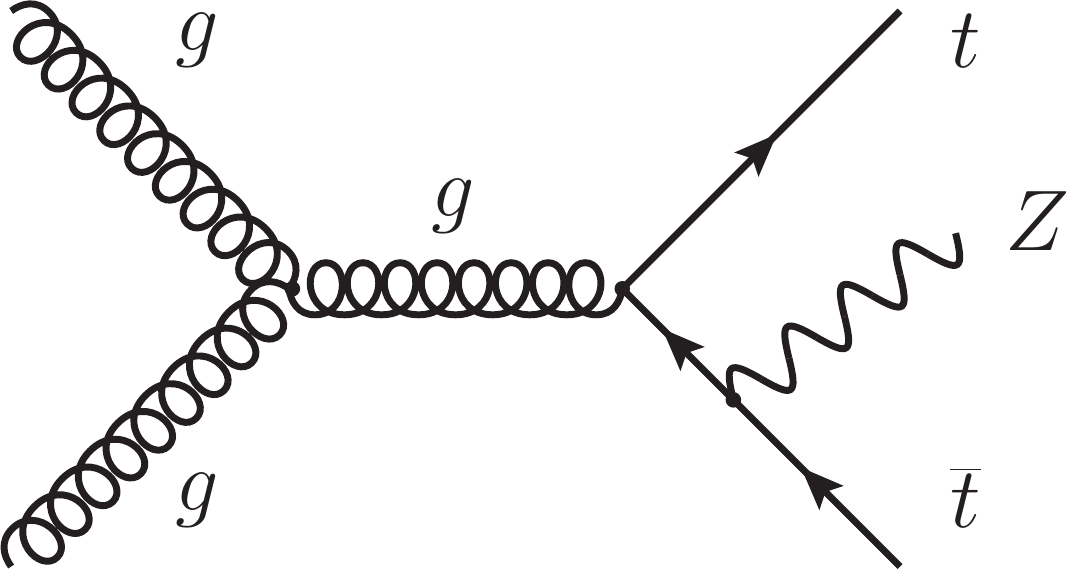}
\caption{The dominant leading-order Feynman diagrams for \ttw and \ttz production in pp collisions.
  The charge conjugate of the diagrams shown is implied.\label{fig:diagrams}}
\end{figure}

Figure~\ref{fig:diagrams} shows the most important leading-order Feynman diagrams for \ttw and \ttz production
in pp collisions. For pp collisions at $\sqrt{s}=8\TeV$, the current best estimates of
the cross sections for these processes are based on quantum chromodynamics~(QCD) calculations at
next-to-leading-order (NLO) in $\alpha_s$. Using CT10~NLO~\cite{ct10} parton distribution functions (PDF) and
a top-quark mass of 173\GeV, the software framework
\MADGRAPH{}5\_a\MCATNLO~\cite{MG5NLO,amcatnlo} provides a cross section of $206^{+21}_{-23}\unit{fb}$ for \ttw production
and of $197^{+22}_{-25}\unit{fb}$ for \ttz production, in
agreement with independent NLO calculations~\cite{ttWxsecNLO,ttZxsecNLO}.

As the number of selected signal events is expected to be comparable to that of the background processes,
the estimation of the background is a key aspect of the analysis. The strategy is to use
background-dominated control samples in data to the maximum extent possible.
Many contributions to the background, in particular those caused by detector
misreconstruction, are estimated in this way, while the remaining irreducible backgrounds
are estimated using Monte Carlo (MC) simulations and the most precise calculations of cross sections that are
available.
For the three separate channels and also for their combination, the yields of events found in excess of the
expected backgrounds are used to measure the corresponding signal cross sections.

\section{The CMS detector\label{sec:detector}}
The central feature of the CMS apparatus is a superconducting solenoid
of 6\unit{m} internal diameter, providing a magnetic field of 3.8\unit{T}. Within the superconducting
solenoid volume are a silicon pixel and strip tracker, a lead tungstate crystal electromagnetic
calorimeter, and a brass/scintillator hadron calorimeter. Muons are measured in
gas-ionization detectors embedded in the steel flux-return yoke outside the solenoid.
A forward calorimeter extends the coverage provided by the barrel and endcap detectors.
CMS uses a right-handed coordinate system, with the origin at the nominal interaction point, the $x$ axis
pointing to the centre of the LHC, the $y$ axis pointing up (perpendicular to the LHC plane), and the $z$
axis along the anticlockwise-beam direction. The polar angle $\theta$ is measured from the positive $z$
axis and the azimuthal angle $\phi$ is measured in the $x$-$y$ plane in radians.
Events are selected by a two-stage trigger system: a hardware-based trigger
followed by a software-based high-level trigger running on the data acquisition computer farm.
A more detailed description of the CMS apparatus can be found in Ref.~\cite{cmsJINST:2008}.
\section{Event selection and Monte Carlo simulation}
\label{sec:evsel}

For all the channels considered in this analysis, the data are selected online by
dilepton ($\Pe\Pe$, $\Pe\Pgm$, and $\Pgm\Pgm$) triggers that demand a transverse momentum (\pt)
larger than 17\GeV for the highest \pt lepton and 8\GeV for the second-highest.
The online selection involves loose identification for both flavours and isolation requirements on
electrons. Other channel-specific triggers, which are described in detail later,
select control regions that are used for the estimation of specific backgrounds and the assessment
of the signal selection efficiency.
After the online selection, data and simulated events are reconstructed offline using the same software.

Each event is processed using a global event reconstruction approach ~\cite{CMS-PAS-PFT-09-001,CMS-PAS-PFT-10-001}.
This consists
in reconstructing and identifying particles using
an optimized combination of the information from all subdetectors.
In this process, the identification of the particle type (photon, electron, muon, charged hadron, and neutral hadron)
plays an important role in the determination of the particle direction and energy.

The tracks reconstructed in the silicon tracker are clustered in several primary vertices corresponding to the
different pp interactions occurring within the same LHC bunch crossing. The vertex that has
the largest $\sum_i p_{\mathrm{T}_i}^{2}$, where $i$ runs over all tracks of the vertex, is assumed to identify
the signal primary vertex. Its position is used to discriminate
against particles originating from the other interactions (pileup) and to distinguish between prompt and non-prompt
particles stemming from the signal interaction.

For each event, hadronic jets are clustered from the reconstructed particles using the
anti-\kt algorithm~\cite{antiKT:2008}, operated with a distance parameter of 0.5. The jet momentum
is determined as the vector sum of all particle momenta in the jet.
In this analysis the jets used for the definition of the signal regions (signal jets) are required to be inside the tracker
acceptance, \ie $\abs{\eta}<2.4$ where $\eta \equiv -\ln [ \tan(\theta/2)]$, to reduce the uncertainty in the jet
reconstruction efficiency and
improve the precision of the energy measurement. Jet energy corrections are applied to
account for the non-linear response of the calorimeters and other instrumental effects. These corrections are based on in
situ measurements using dijet and $\gamma+\text{jet}$ data samples~\cite{JES}.
A two-fold approach is employed to reduce the effect of pileup during jet reconstruction. Firstly,
charged particles whose trajectories
point to pileup vertices are excluded from the set of particles that are used for the reconstruction of signal jets.
Secondly, the average energy density due to neutral pileup particles is evaluated in each event, and the
corresponding energy inside the jet is subtracted~\cite{Cacciari:2008gn}.
Then a jet identification requirement~\cite{generalJetId},
primarily based on the energy balance between charged and neutral hadrons in a jet, is
applied to remove jets that are misreconstructed or originate from instrumental noise.
Finally, the trajectories of all the charged particles of a jet are
used to calculate a \pt-averaged longitudinal impact parameter for each signal jet~\cite{jetIDPU}. This variable is then
employed as a discriminator against jets from pileup.
Unless otherwise specified, signal jets are required to have $\pt > 30$\GeV.

To identify (tag) jets originating from the hadronization of bottom quarks (b jets), the combined secondary vertex (CSV)
algorithm~\cite{btagging:8TevPAS} is used. The algorithm combines the information about track impact parameters
and secondary vertices within jets in a likelihood discriminant to provide
separation between b jets and jets originating from light quarks, gluons, or charm quarks. We use
here two operating points. The \emph{loose} working point corresponds to
a b-tagging efficiency for jets originating from $\cPqb$ quarks of about 85\% and a
misidentification probability for jets from light quarks and gluons of 10\%.
The \emph{medium} working point provides an efficiency
of about 70\%  and a misidentification probability of 1.5\%.

Muons and electrons are identified using standard quality criteria~\cite{MUOPAS,EGMPAS} and
are required to have $\pt > 20$\GeV and $\abs{\eta}<2.4$. For the four-lepton
channel only, identified leptons with \pt between 10 and 20\GeV are also employed for the event selection.
To reduce the contamination caused by leptons from heavy-flavour decays or misidentified hadrons in
jets, leptons are required to be isolated and to pass a selection on the impact parameter, which is calculated with
respect to the position of the signal primary vertex. Candidates are considered
isolated when the ratio of the scalar sum of the transverse momenta of all the other reconstructed particles in a cone
of $\Delta R = \sqrt{\smash[b]{(\Delta\eta)^2 + (\Delta\phi)^2}} = 0.3$ around the candidate, relative to the
lepton \pt value, is less than 5--10\%, the exact value of the threshold depending on the flavour of
the lepton and on the final state.
This relative isolation is corrected for the expected contribution from
pileup using an approach that is similar to the one employed for the reconstruction of jets~\cite{HiggsObservationCMS}.
The leptons are required to originate from the primary interaction
demanding that their transverse and longitudinal impact parameters are smaller than 50--200\micron and 0.1--1.0\cm,
respectively. The tightest selections in these ranges are used for the lepton flavour and final states that are most affected by
backgrounds due to non-prompt leptons.

Finally, the observables \MET and \HT are used, respectively, to identify the presence of neutrinos and
to measure the hadronic activity in the analysed events. The former is defined as the magnitude of the vector sum
of the transverse momenta of all reconstructed particles, the latter is the scalar sum of the transverse momenta
of all signal jets.

Simulations, which include pileup effects, are used to estimate some of the backgrounds, as well as to
calculate the selection efficiency for the \ttw and \ttz signal events.
Simulated samples are generated with the \MADGRAPH{}5~\cite{MADGRAPH5} program, with the
exception of the \ttH background process that is generated using \PYTHIA~6~\cite{PYTHIA}.
All simulated samples are processed using a \GEANTfour-based model~\cite{Geant} of the CMS detector. Signal
samples are produced with \MADGRAPH{}5, which is used with the CTEQ6L1~\cite{cteq} PDF and is interfaced to
\PYTHIA~6.424 to simulate parton showering and hadronization.

\section{Same-sign dilepton analysis} \label{sec:ssdl}
\label{sec:ssdl_intro}

The aim of the same-sign dilepton analysis is to search for \ttw events where one lepton is produced in the leptonic decay chain of one of the two top quarks, and the other like-sign
lepton stems directly from the decay of the prompt vector boson:
\begin{equation*}
\Pp\Pp\to \ttw \to  (\cPqt\to \cPqb\ell\nu)(\cPqt\to \cPqb\cPq\cPaq')(\W \to \ell\nu),
\end{equation*}
where $\ell$ corresponds to an electron or a muon.
By requiring that the two selected leptons
have the same sign, only half of the signal produced
in the dilepton final state can be selected. However, the requirement significantly improves the
signal-over-background ratio.
The main background
is caused by misidentification and misreconstruction effects:  decay products of heavy-flavour
mesons that give rise to non-prompt leptons and pions in jets
misidentified as prompt leptons. A second, smaller,
source of background is also caused by misreconstruction and consists of opposite-sign dilepton events where the charge
of one of the two leptons is wrongly assigned.

The selection for the dilepton channel is conducted through the following steps:
\begin{enumerate}
	\item \label{cutLeps} Each event must contain two isolated leptons of the same charge and $\pt > 40\GeV$.
        Both leptons are required to be compatible
        with the signal primary vertex and have a relative isolation smaller than $5\%$.
        The invariant mass of the dilepton pair is required to be larger than 8\GeV.
	\item \label{cutJets} Three or more signal jets must be reconstructed, and at least one of these has to be b-tagged using
        the medium working point of the CSV algorithm.
        \item \label{cutLepVeto} Events are rejected if they contain a third lepton forming, with one of the other two
        leptons, a same-flavour opposite-sign pair whose invariant mass is within 15\GeV of the
        known Z-boson mass~\cite{Beringer:1900zz}. For the third lepton, the relative isolation must be less than $ 9\:(10)$\%
        if it is an electron (muon), and the transverse momentum requirement is loosened to $\pt > 10$\GeV.
	\item \label{cutHT} The \HT value is required to be greater than 155\GeV.
        \item \label{cutCharge} Selected events are grouped in three categories depending on the lepton flavour: $\Pe\Pe$, $\Pe\Pgm$, and $\Pgm\Pgm$ dilepton
        pairs. Each of these categories is further split into two separate sets of dileptons with either positive or negative charges,
        for a total of six signal regions.
\end{enumerate}

The tight-lepton selection~(\ref{cutLeps}) reduces the background from misidentified leptons, while the invariant
mass requirement rejects events with pairs of energetic leptons from decays of heavy hadrons.
The requirement~(\ref{cutJets}) on the general number of jets and on the number of b-tagged jets present in the event decreases
the background from electroweak processes, \eg WZ production, that can have same-sign leptons in the final state, but
are accompanied by little hadronic activity. The WZ background is also significantly reduced by the third-lepton veto (\ref{cutLepVeto}).
The \HT requirement~(\ref{cutHT}) as well as the threshold on the lepton \pt (\ref{cutLeps}) have been optimized
for the best signal significance. This selection also minimizes the expected uncertainty in the measured cross section.
The splitting~(\ref{cutCharge}) of the signal candidates in six categories  is done for two reasons: exploiting the smaller
background from lepton and charge misidentification in signal regions with muons and benefitting
from the greater signal cross section in the plus-plus dilepton final states, which is caused by the abundance of quarks, instead of
antiquarks, within the colliding protons at the LHC.
Finally, the Z-boson veto is necessary to make the dilepton analysis statistically independent from the trilepton one described later.
Events with three leptons are not rejected if they pass the Z-boson veto, since these can stem from a fully-leptonic decay
of the \ttbar pair in \ttw signal events.

\subsection{Background estimation} \label{sec:2L_bkg}

{\tolerance=800
After the full same-sign dilepton selection is applied, there are three general categories of background processes
that are selected together with \ttw signal events: background from non-prompt or
misidentified leptons (\emph{misidentified lepton} background); background from lepton charge
misidentification (\emph{mismeasured charge} background); WZ and \ttz
production, as well as other rare SM processes that contain genuine pairs of prompt, isolated and same-sign leptons.
The subset of these processes that do not contain a Z boson in the final state forms the {\em irreducible}
component of the background. This includes the production of like-sign WW
and the production of the Higgs boson in association with a pair of top quarks.
The production of a \ttbar pair in association with a W boson by means of double parton scattering
is expected to have a cross section two orders of magnitude smaller than the \ttw production through single scattering~\cite{DPS}.
This source of background is therefore considered negligible and is ignored in the rest of the analysis.
\par}

{\tolerance=600
The first background consists mostly of \ttbar events, with a second important contribution coming
from \wjets events. In both cases, one prompt lepton originates from the leptonic decay of a W boson,
and another same-sign lepton is caused by the misidentification of a non-prompt lepton
stemming from the decay of a heavy-flavour hadron.
In \wjets events, smaller sources of misreconstructed leptons affecting this category of background are given by
the misreconstruction of hadrons, the production of muons from light-meson decays, and the reconstruction of
electrons from unidentified photon conversions.
The background yield is estimated from data using a sample of events that satisfy the full analysis selection,
except that one of the two leptons is required to pass a looser lepton selection and fail the full selection
(sideband region). The background rate is then obtained weighting the events in this sideband region by the
``tight-to-loose'' ratio, \ie the probability for a loosely identified lepton to pass the full
set of requirements.
This tight-to-loose ratio is measured as a function of lepton \pt and $\eta$
in a control sample of dijet events, which is
depleted of prompt
leptons and is selected by dedicated
single-muon and single-electron triggers.
The systematic uncertainty in the background estimate is due to the differences in the
various sources of non-prompt or misidentified leptons, between the dijet events where the tight-to-loose
ratio is measured and the sideband region where the ratio is applied. Among the most important differences
are the \pt spectrum and the flavour of the jets containing the misidentified leptons.
These two quantities have been varied in the control sample using appropriate selections and then the effects
on the tight-to-loose ratio, and on the background estimate itself, have been quantified. The
range of variation for these two quantities has been guided by a simulation of the background processes.
The full systematic uncertainty in the background is estimated to be 50\%.
The statistical part of the uncertainty is driven by the number of events in the sideband region
and it is significantly smaller than the systematic uncertainty for all six signal regions.
\par}

The probability to misidentify the charge of muons is about an order of magnitude smaller than for electrons.
Therefore the magnitude of the background caused by charge misidentification, mostly in Drell--Yan and \ttbar events, is
driven only by electrons. This
background is estimated by
selecting opposite-sign \Pe\Pe\ or \Pe\Pgm\
events that pass the full analysis selection, except the same-sign requirement, and then weighting them by the
\pt- and $\eta$-dependent probability for electron charge
misassignment. This probability and its variation as a function of the
lepton \pt and $\eta$ are determined by combining information from simulation and a control data sample
of $\cPZ \to \Pe \Pe$ events.
For the electron selection used in this analysis, the probability of charge
misidentification is about $10^{-4}$ and $10^{-3}$ for electrons reconstructed
in the barrel and endcap detectors, respectively. The background estimate has an uncertainty
of 30\% (15\%) for the \Pe\Pe\ (\Pe \Pgm) signal regions.
This uncertainty accounts for
differences between data and simulation, and the limited momentum range of electrons
in the $\cPZ$-boson control sample.

Production of WZ and \ttz events, and the irreducible backgrounds, are all
estimated from simulation
as done when calculating the signal selection efficiencies.
For each SM process contributing to this category of background, the dominant systematic uncertainty
is the one in the theoretical cross section prediction. Depending on the process, we use an uncertainty of 15--50\%
and consider it as fully correlated across all signal regions.

\subsection{Same-sign dilepton results}	\label{sec:resultsSS}

After the full analysis selection is applied, 36 events are observed in data, to be compared with
$25.2\pm 3.4\systOstat$ events expected from background processes and $39.7\pm 3.5\systOstat$
events from the sum of background and \ttw signal with the SM cross section.
For both predictions, the statistical and systematic uncertainties are added in quadrature.

\begin{figure*}[htbp]
 	\centering
	\includegraphics[width=0.45\textwidth]{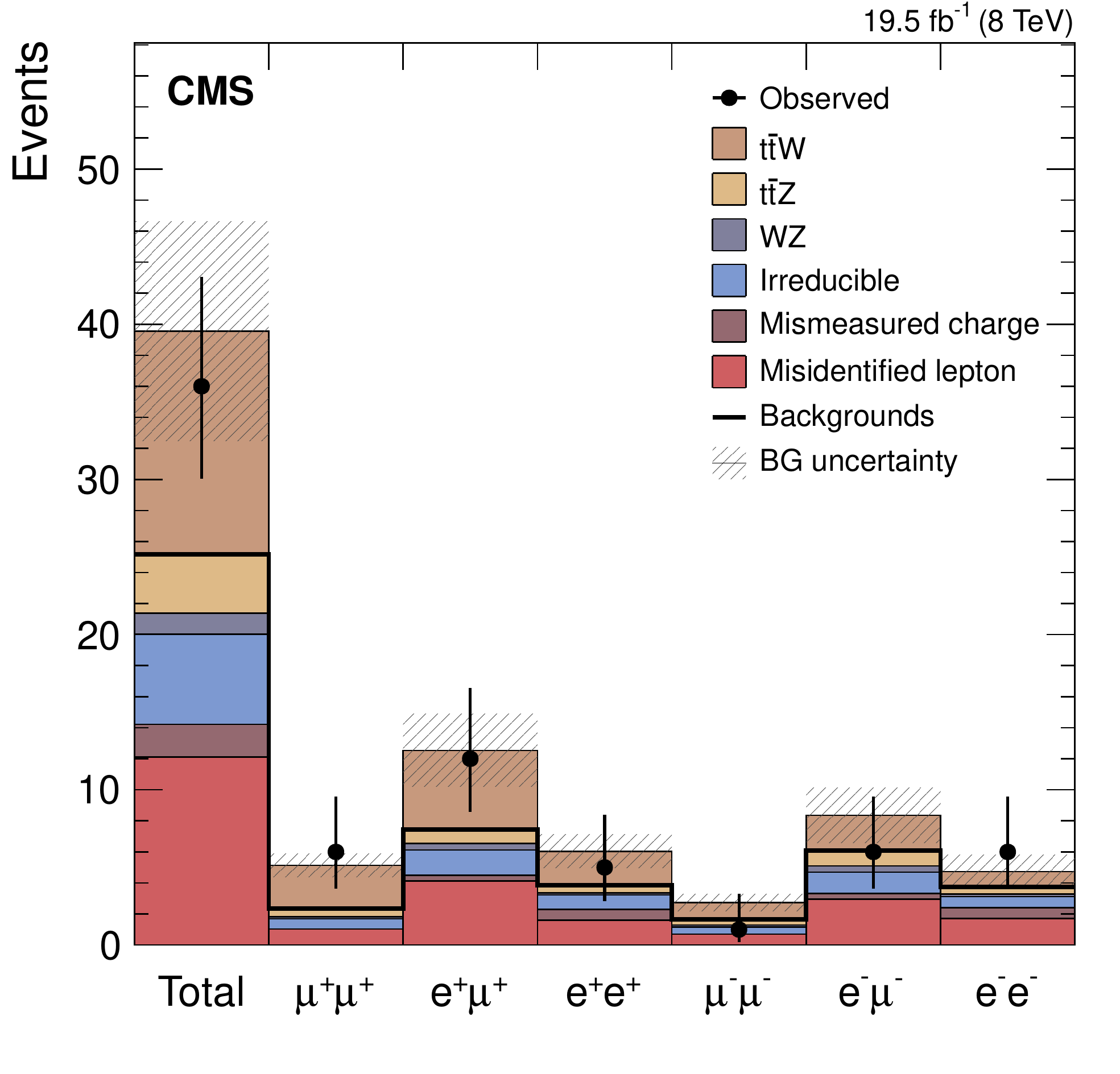}
	\includegraphics[width=0.45\textwidth]{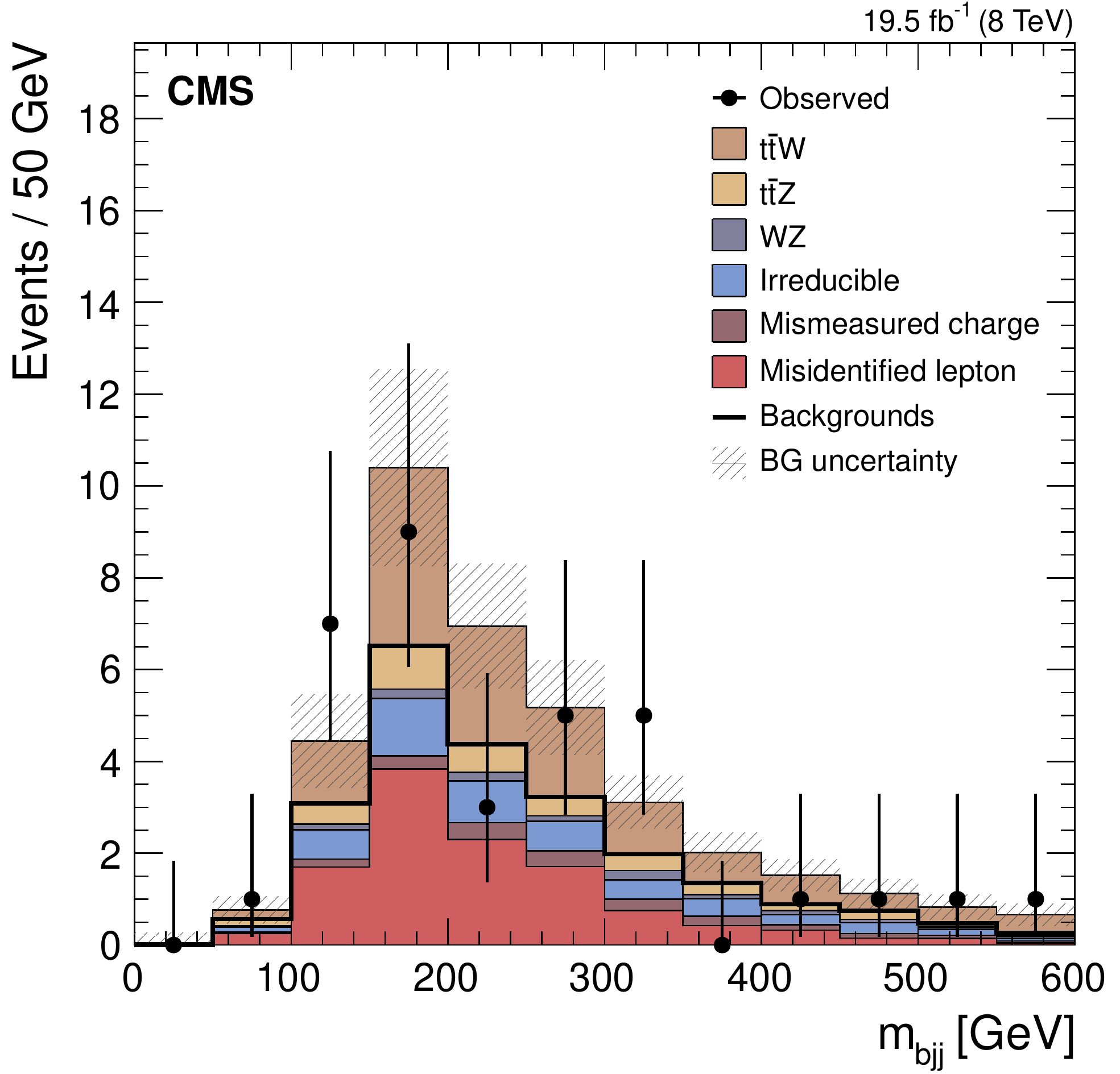}
	\includegraphics[width=0.45\textwidth]{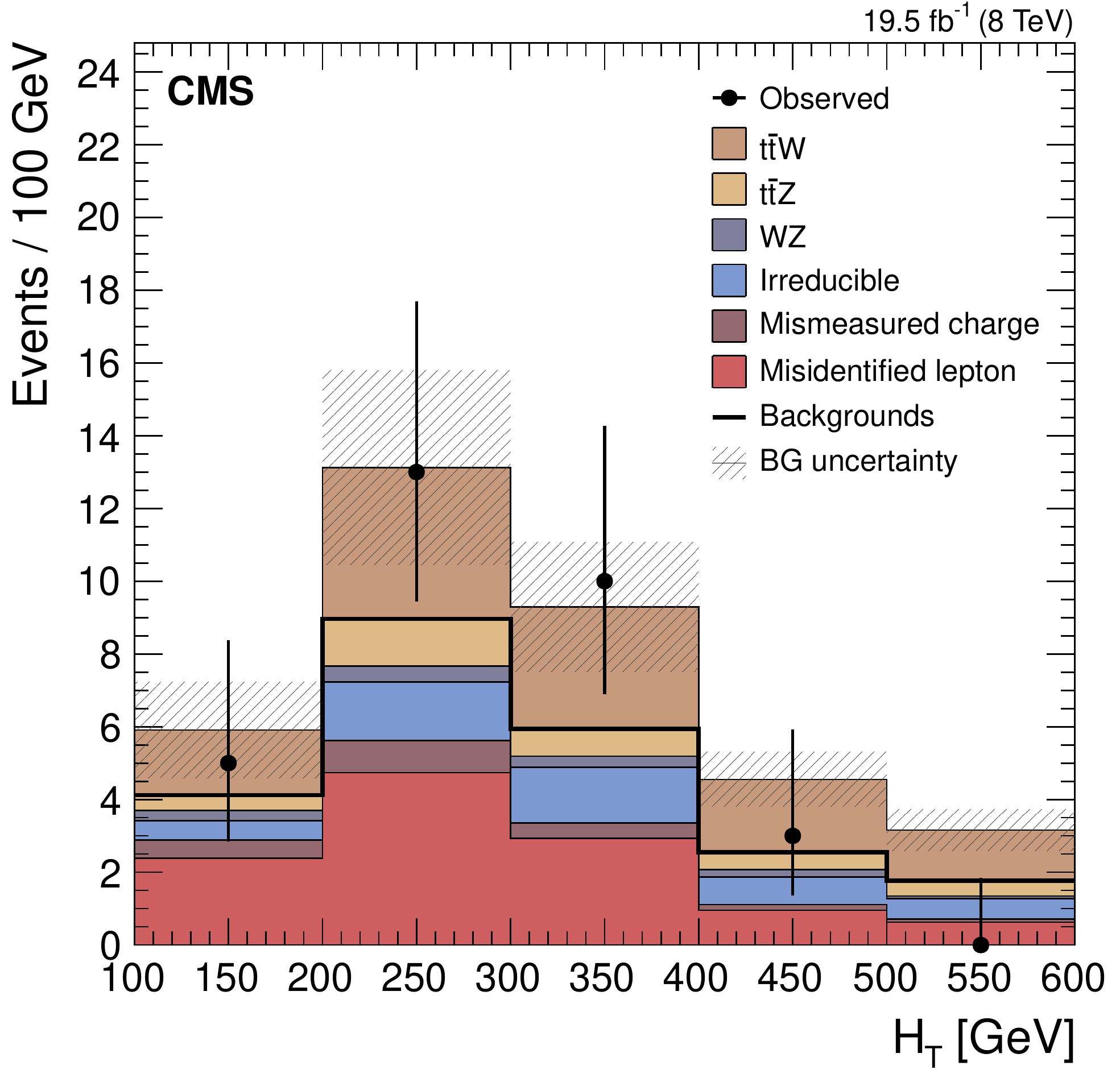}
	\includegraphics[width=0.45\textwidth]{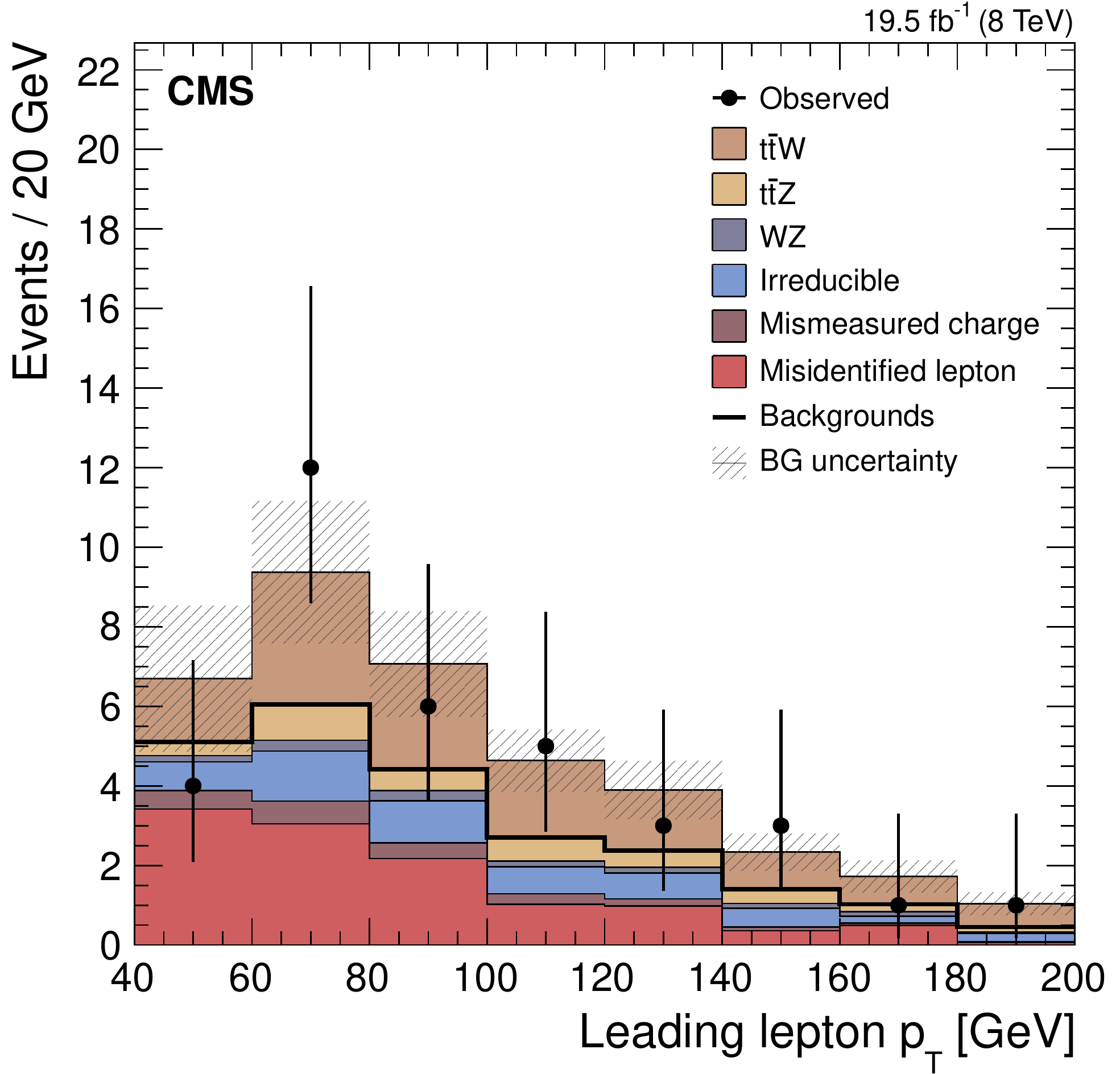}
	\caption{Event yields in data after final dilepton selection requirements, compared to the background estimates
        and signal expectations.
        Contributions separated by final states (top left), trijets mass distribution for the hadronic top-quark candidate (top right),
        \HT distribution (bottom left), and the leading-lepton \pt distribution (bottom right).
        The combination of statistical and systematic uncertainties is denoted by the shaded area.}

	\label{fig:yieldSS}
\end{figure*}

The event yields, along with
the corresponding uncertainties for each background component, are reported in Table~\ref{tab:yieldSS}.
The top left panel of Fig.~\ref{fig:yieldSS} shows the distribution of
the expected and observed events across the six different signal regions, and for all dilepton channels added together.
As already anticipated, the positively charged
channels are expected to collect a larger quantity of signal than the negatively charged channels,
for a comparable quantity of background.  The first three channels therefore drive the sensitivity of this analysis.
In Table~\ref{tab:yieldSS} and Fig.~\ref{fig:yieldSS}, and in the equivalent tables and figures for the other
two leptonic channels, the uncertainty in the signal cross section is not shown because
it does not affect the precision of the experimental measurement.

\begin{table*}[htb]
\centering
\topcaption[]{Expected signal, estimated backgrounds, the sum of the two, and observed number of events for the $\mu^{\pm}\mu^{\pm}$,
  $\Pepm\mu^{\pm}$, and $\Pepm\Pepm$ channels. Uncertainties include both the statistical and the systematic components.
  The systematic uncertainty in the signal contribution does not include
  the theoretical uncertainty in the signal production cross section.}
\label{tab:yieldSS}

\begin{tabular}{lr@{$\,\pm\,$}lr@{$\,\pm\,$}lr@{$\,\pm\,$}lr@{$\,\pm\,$}lr@{$\,\pm\,$}lr@{$\,\pm\,$}l}
\hline
                  &\multicolumn{2}{c}{$\Pgmp\Pgmp$}&\multicolumn{2}{c}{$\Pep\Pgmp$}&\multicolumn{2}{c}{$\Pep\Pep$} &\multicolumn{2}{c}{$\Pgmm\Pgmm$}&\multicolumn{2}{c}{$\Pem\Pgmm$}&\multicolumn{2}{c}{$\Pem\Pem$} \\
\hline
\ttw (expected) &           2.8 &           0.4 &           5.1 &           0.5 &           2.2 &           0.3 &           1.1 &           0.2 &           2.3 &           0.3 &           1.0 &           0.2 \\
\hline
Misidentified lepton &           1.0 &           0.6 &           4.1 &           2.1 &           1.6 &           0.9 &           0.7 &           0.4 &           3.0 &           1.5 &           1.7 &           0.9 \\
 Mismeasured charge & \multicolumn{2}{c}{----}        &           0.4 &           0.1 &           0.7 &           0.2 & \multicolumn{2}{c}{---}        &           0.4 &           0.1 &           0.7 &           0.2 \\
        Irreducible &           0.7 &           0.4 &           1.6 &           0.9 &           0.9 &           0.5 &           0.5 &           0.3 &           1.4 &           0.7 &           0.7 &           0.4 \\
                 WZ &           0.1 &           0.1 &           0.4 &           0.1 &           0.1 &           0.1 &           0.1 &           0.1 &           0.4 &           0.1 &           0.2 &           0.1 \\
               \ttz &           0.6 &           0.3 &           0.9 &           0.5 &           0.5 &           0.3 &           0.4 &           0.2 &           1.0 &           0.5 &           0.5 &           0.3 \\
\hline
        Total background &           2.4 &           0.7 &           7.4 &           2.3 &           3.9 &           1.1 &           1.7 &           0.5 &           6.1 &           1.8 &           3.7 &           1.1 \\
    Total expected &           5.2 &           0.8 &           12.5 &           2.4 &           6.1 &           1.1 &           2.8 &           0.5 &           8.4 &           1.8 &           4.7 &           1.1 \\
\hline
{ Observed} &  \multicolumn{2}{c}{6} &  \multicolumn{2}{c}{12} &  \multicolumn{2}{c}{5} &  \multicolumn{2}{c}{1} &  \multicolumn{2}{c}{6} &  \multicolumn{2}{c}{6}  \\
\hline

\end{tabular}
\end{table*}

{\tolerance=600
The other three panels of Fig.~\ref{fig:yieldSS} show the distributions for the invariant mass $m_{\mathrm{bjj}}$ of the
three jets expected to originate from the hadronic top-quark decay (top right), \HT (bottom left), and the
leading-lepton \pt (bottom right) for all six signal regions combined together.
For each event, the three signal jets used for the $m_{\mathrm{bjj}}$ distribution are selected as follows:
one, and only one, of the three jets is b-tagged; among the possible three-jet combinations the one chosen minimizes
$\Delta R_{\mathrm{jjj}} = \sqrt{\smash[b]{ (\Delta R_{\mathrm{j}_1,\cPqt})^2 + (\Delta R_{\mathrm{j}_2,\cPqt})^2 + (\Delta R_{\mathrm{j}_3,\cPqt})^2 }}$,
where $\Delta R_{\mathrm{j}_i,\cPqt}$ is the $\Delta R$ distance between the direction of the $i$-th jet and the
direction of the reconstructed hadronic top-quark candidate. In all four distributions data and simulation are found in agreement.
In particular, the $m_{\mathrm{bjj}}$ distribution confirms that most of the background from misidentified
leptons is originating from top-quark events. In Fig.~\ref{fig:yieldSS}, and also in all other similar
figures included in this document, the error bar enclosing each data point represents the 68\% confidence level
interval around the mean of the corresponding Poisson distribution.
\par}

Based on the observed number of events, the background estimates, and the signal acceptance
(including the leptonic branching fractions), the inclusive \ttw production cross
section is measured, through the combination of the six dilepton channels, as
\begin{equation*}
\sigma_{\ttw} = 170  ^{+90}_{-80}\stat \pm 70\syst\unit{fb},
\end{equation*}
including statistical and systematic uncertainties, compared to the SM expectation of  $206^{+21}_{-23}\unit{fb}$.
The significance of the result over the background-only hypothesis is equivalent to 1.6 standard
deviations (2.0 standard deviations expected).

The systematic uncertainty in the signal selection efficiency is 8\%. It is treated in a common way
with the three- and four-lepton channels and is discussed in detail in Section~\ref{sec:signalSyst}.
Additionally, for all channels there is a 2.6\% uncertainty in the expected yield of signal and simulation-derived
background events because of the uncertainty in the luminosity
normalization~\cite{CMS-PAS-LUM-13-001}.
However, together with the low yield of signal events, the main factor dominating the uncertainty in the cross section
measurement is the uncertainty in the largest background
component, \ie the 50\% uncertainty in the background from misidentified leptons.

\section{Trilepton analysis} \label{sec:trilepton}
The production of a \ttbar pair in association with a \cPZ~boson is analysed in the final state
with three high-energy, isolated, and prompt leptons.
The trilepton analysis targets final states with only one \PW~boson decaying leptonically:
\begin{equation*}
\Pp\Pp\to \ttz \to  (\cPqt \to \cPqb \ell\nu)(\cPqt \to \cPqb \cPq\cPaq')(\cPZ \to \ell\overline{\ell}).
\end{equation*}
The event selection, described in more detail below, focuses on the main features of this final state:
two oppositely charged leptons of the same flavour, consistent with the \cPZ-boson decay;
an additional lepton; and at least four jets, at least two of which  are b-tagged.
The isolation of the leptons has additionally been loosened to reflect the diminished contribution of misidentified leptons to the background.

The selection for the trilepton channel is conducted through the following steps:
\begin{enumerate}
   \item \label{3L:cutLeps} Each event must contain three isolated leptons of $\pt > 20\GeV$ and passing 
        identification requirements described in Section~\ref{sec:evsel}. All three leptons are required to be compatible
        with the signal primary vertex and have a relative isolation smaller than~9\%~(10\%) for electrons (muons).
    \item \label{3L:cutZWindow} Two of the leptons must be of the same flavour, be oppositely charged,
          and form  an invariant mass between 81 and 101\GeV to be consistent with a \cPZ-boson decay. If
          multiple pairs pass this selection, the one with the mass closest to the known \cPZ-boson mass
          is selected as the \cPZ~boson candidate.
    \item \label{3L:cutJets} To match the final-state signal topology, four or more signal jets
          must be reconstructed with at least three of these jets having $\pt>30\GeV$, and the
          fourth jet is required to have $\pt>15\GeV$. Additional identification and pileup suppression
          selections are applied as described in Section~\ref{sec:evsel}.
    \item \label{3L:cutbTags}     %At least two of the jets must be b-tagged and have $\pt>30\GeV$, the first using
          At least two of the jets with $\pt>30\GeV$ must be b-tagged, the first using
          the medium working point of the CSV algorithm, and the second using the loose working point.
    \item \label{3L:cutLepVeto} Events are rejected if they contain a fourth lepton with a
          loosened transverse momentum requirement of $\pt > 10$\GeV, in order not to overlap with
          the four-lepton analysis.
\end{enumerate}

These event selections have been optimized for the best precision on the expected measured cross section.
A broad range of variations to the applied requirements has been considered in the optimization:
including in the event selections a minimum number of jets, minimum jet \pt, as well as \HT;
changing the number of jets required to be b-tagged; and varying the lepton momentum and isolation thresholds.
Estimates of the expected backgrounds used in the optimization of the final requirements
have been made both with initial estimates from simulation alone as well as with events in data control samples
using the methods described below.

\subsection{Background estimation\label{sec:trilep_bgd}}
Backgrounds passing the analysis selections are separated into three components:
irreducible contributions from events with three prompt leptons and two b-quark jets
({\em irreducible} component), primarily with at least one top quark in the process;
those with three prompt leptons and b-tagged jets without top-quark contributions
({\em non-top-quark} component);
and contributions with at least one misidentified lepton
({\em misidentified lepton} component).
This categorization is driven by the choice of methods used to estimate the backgrounds.

The irreducible component is split evenly among single-top-quark production in association
with a \cPZ~boson (\cPqt\cPqb\cPZ), \ttH, and  \ttw\ production;
additional contributions from production of three bosons and \ttbar associated
with an isolated photon or two additional vector bosons are much smaller, but are still considered.
Since the \ttw\ contribution is constrained by measurements in other (primarily the same-sign dilepton)
final states, its expected SM
contribution of $0.2\pm 0.1\stat$ events is quoted separately.
The remaining irreducible background contributions are estimated directly from simulation:
$0.77\pm0.04\stat\pm0.39\syst$ events are expected.
The systematic uncertainty in this background is conservatively estimated to be 50\%,
dominated by the uncertainty in the cross section, in accordance with corresponding values
used in Section~\ref{sec:2L_bkg}.
This systematic uncertainty is applied also to the \ttw\ contribution and serves
as an initial constraint to the combined measurement, as discussed in Section~\ref{sec:combination}.

The non-top-quark component contributions are primarily from events with three prompt leptons
and b-tagged jets from misidentified light-flavour jets or b-quark jets arising
from initial- or final-state radiation.
In simulation, this contribution is dominated by \PW\cPZ~events.
Because neither the absolute rate of extra jet production from
radiation and higher-order diagrams, nor the flavour composition of
additional jets are well simulated~\cite{BCorrelation}, we rely on data to predict this background.

A sideband sample with three leptons and no b-tagged jets, with all other selections applied,
is dominated by non-top-quark backgrounds and is used to normalize the non-top-quark component prediction.
The method to predict the non-top-quark backgrounds relies on the ratio $R_b$ of the number of events
passing the analysis b-tagging requirements relative to those not having b-tagged jets. This ratio is assumed
to be the same as for inclusive \cPZ+jets production (with the \cPZ~boson
decaying leptonically) for events passing the same jet selections.
We derive the $R_b$ in a sample of events with opposite-sign same-flavour
leptons passing the same identification requirements as in the trilepton sample.
The contribution of \ttbar and other flavour-symmetric backgrounds is subtracted
using opposite-flavour dilepton events after a correction for a difference in the lepton
selection efficiency.
For the final prediction of the non-top-quark component, an additional correction
$C_b=1.4\pm0.2\stat$
is applied based on the difference between the prediction and observation in simulation.
This is done to account for residual differences in the kinematic properties of jets
between \cPZ+jets events and the trilepton non-top-quark background.
The $R_b$ measured in dilepton events in data is $0.160\pm0.003\stat$.
The non-top-quark component is predicted to contribute $2.3\pm0.5\stat\pm1.1\syst$ events.
The systematic uncertainty of approximately 50\%  is estimated as a combination of observed
difference of $R_b$ in the dilepton events between data and simulation
and the deviation of $C_b$ from unity.

Finally, the misidentified-lepton background component is estimated with a method similar to that
of the same-sign dilepton analysis, described in Section~\ref{sec:2L_bkg}.
In each of the four final states the control sample is culled from events
passing the trilepton signal event selections except that only one of the leptons is required
to fail the isolation and identification requirements, still passing looser requirements.
Similar to the same-sign dilepton analysis, the ratio of misidentified
leptons passing full identification and isolation selections relative to
the loosened requirements (the tight-to-loose ratio)
is modelled to be the same in the trilepton events as in a sample with one lepton candidate and a jet.
The modelling is tested in simulation, where the tight-to-loose ratio is measured in simulated multijet
events and is then applied to the dominant background sample, \ie \ttbar production.
The level of agreement between predicted and observed background in simulation gives the leading source
of systematic uncertainty in the method, estimated to be roughly 50\%.
Combined in all trilepton final states, the misidentified lepton component is estimated to be
$1.2\pm 0.5\stat\pm 0.6\syst$ events.

\subsection{Trilepton results}
\label{sec:trilepton_results}

\begin{figure*}[htb]
\centering
\includegraphics[width=0.47\textwidth]{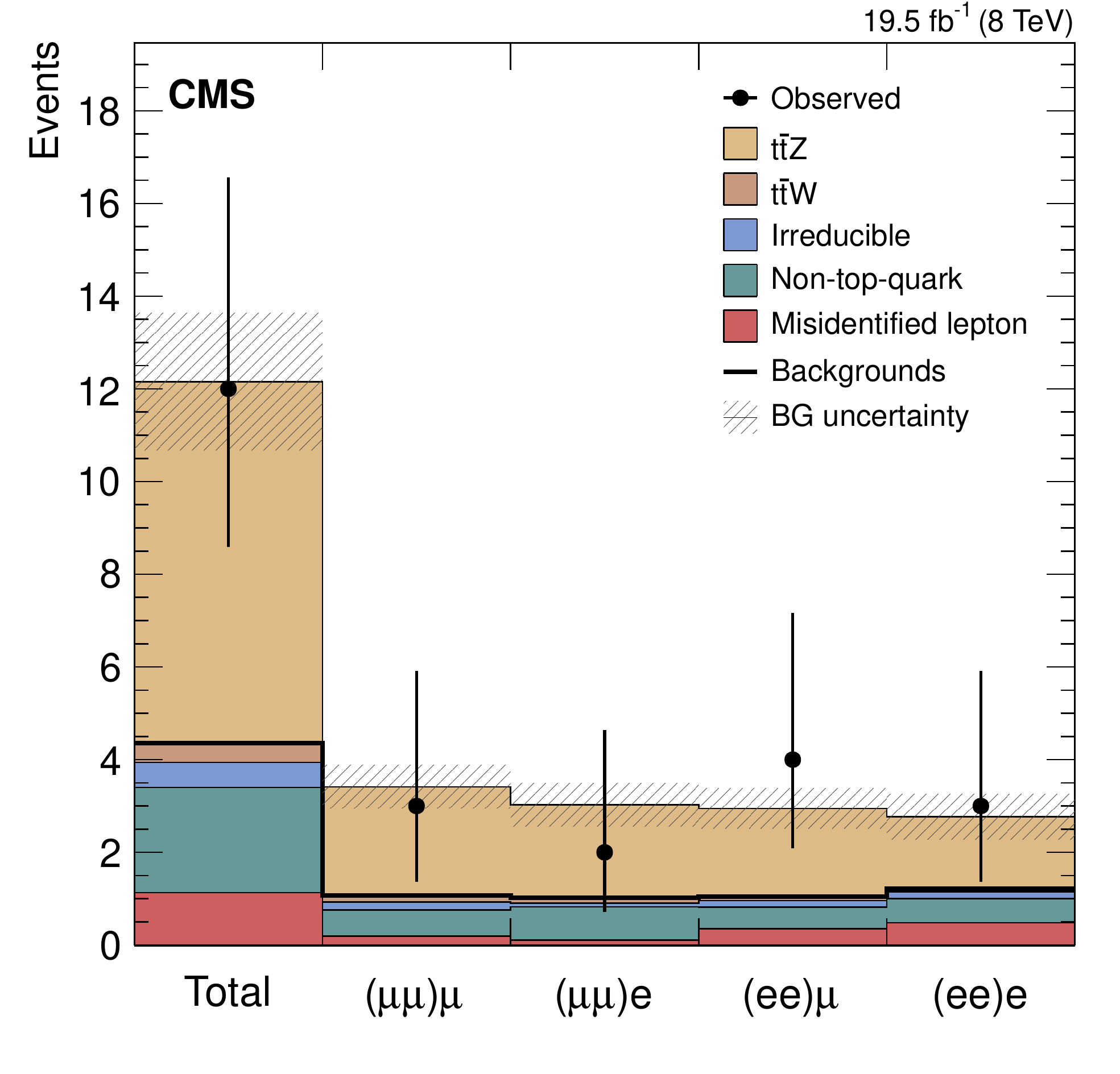}
\includegraphics[width=0.47\textwidth]{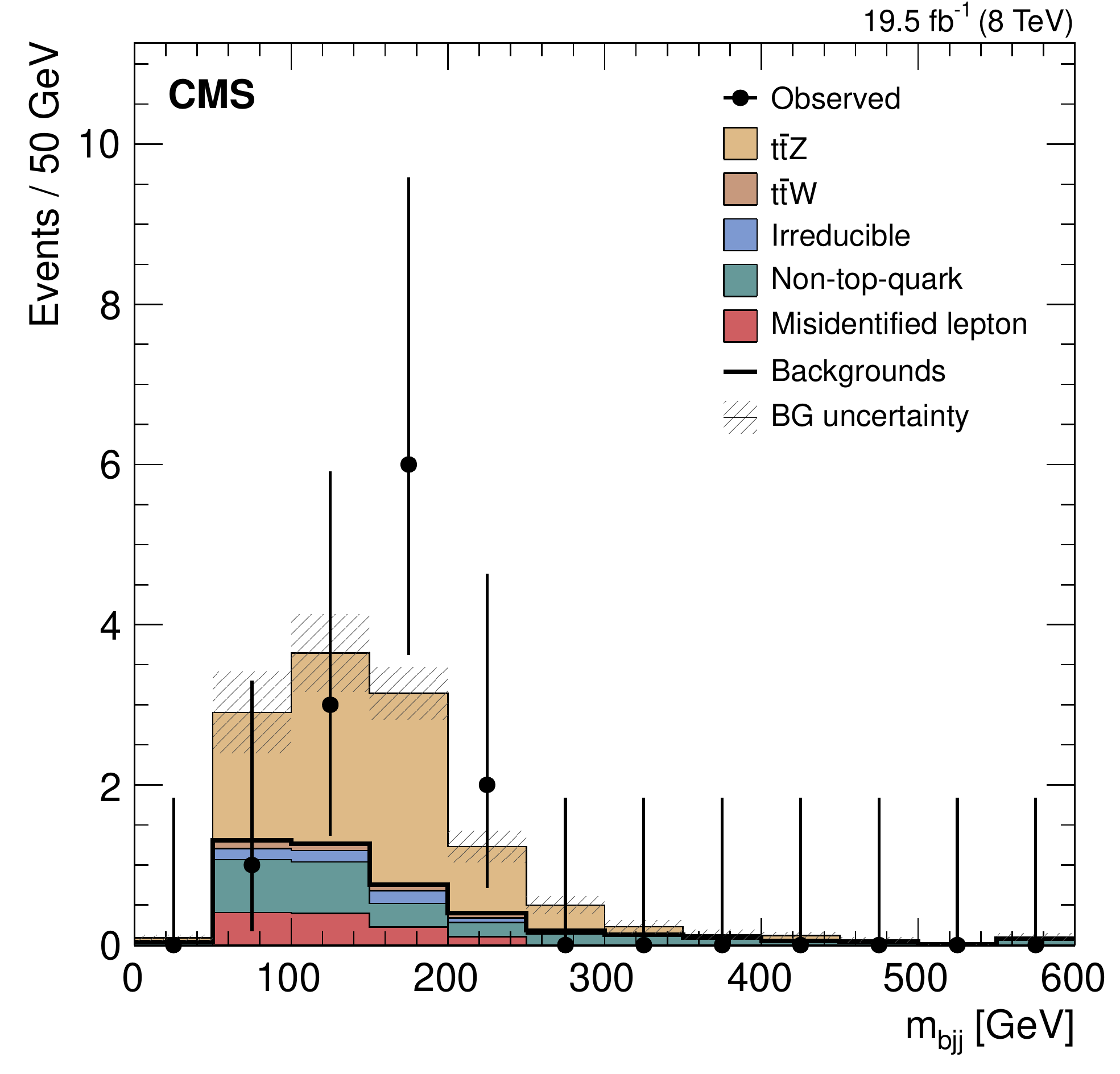}
\includegraphics[width=0.47\textwidth]{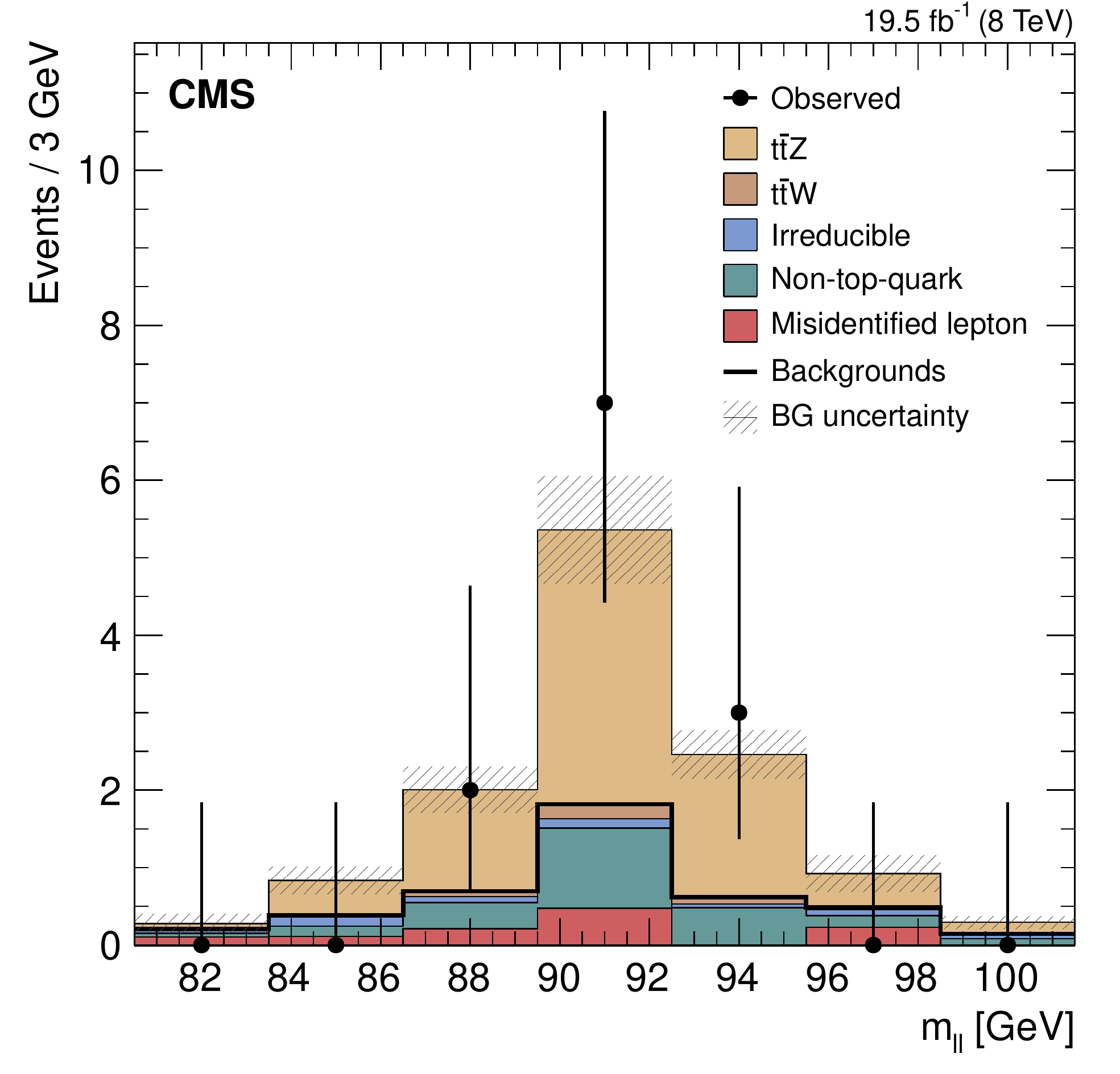}
\includegraphics[width=0.47\textwidth]{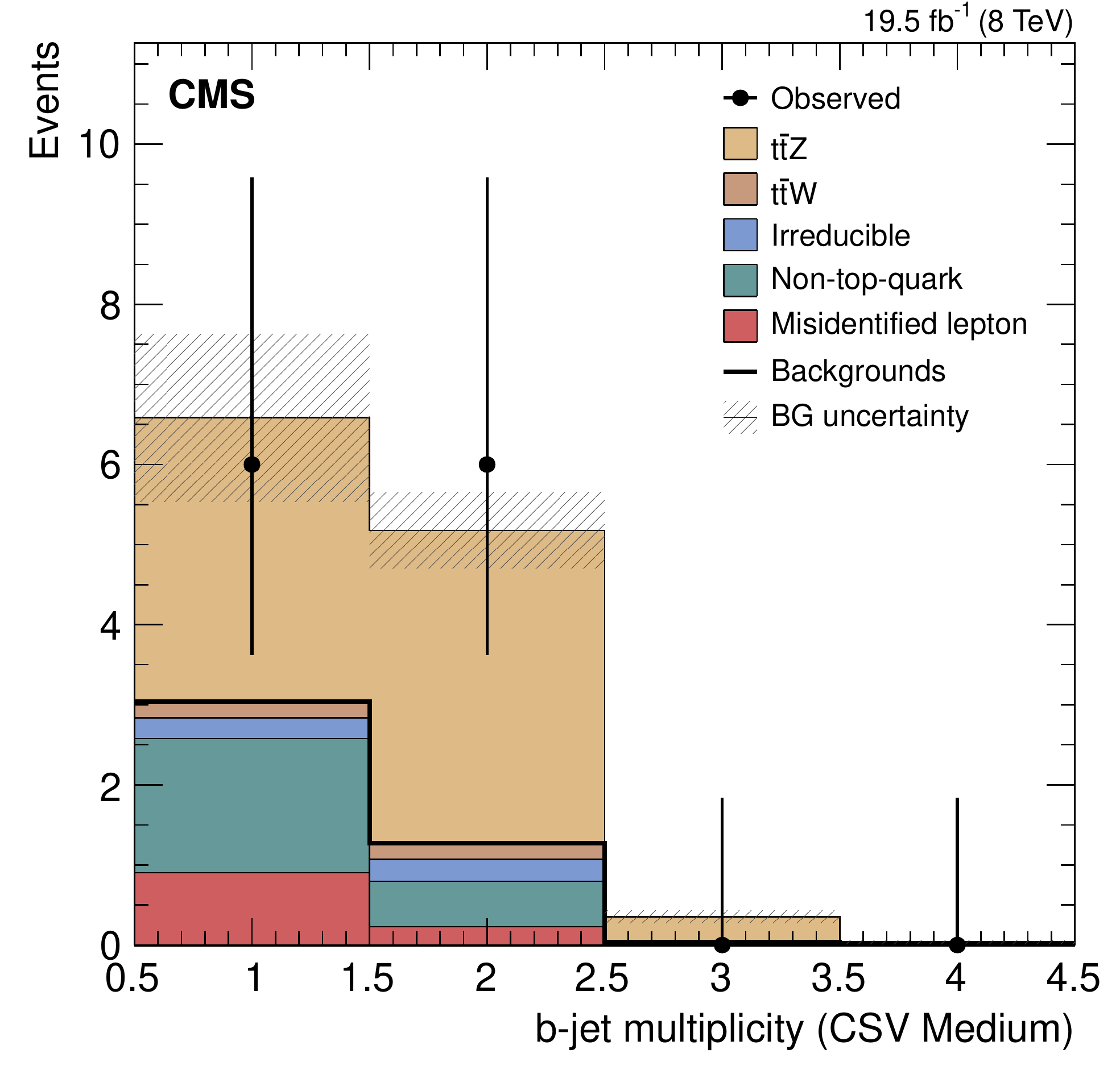}

\caption{Event yields in data after final trilepton selection requirements, compared to the background estimates
  and signal expectations.
  Contributions separated by final states where the two leptons consistent with the Z boson are indicated
  inside parenthesis on the bin labels (top left),
  trijets mass distribution for the hadronic top-quark candidate (top right),
  \cPZ-boson candidate dilepton mass distribution (bottom left),
  and the distribution of the number of b-tagged jets passing medium operating point of the b-tagger (bottom right).
  The combination of statistical and systematic uncertainties is denoted by the shaded area.
 \label{fig:yieldsLLL}}
\end{figure*}

The 12 events observed in data are consistent with the sum of the estimated
backgrounds, $4.4\pm1.6\systOstat$ events, and the expected signal,
$7.8\pm0.9\systOstat$ events.
These results are summarized in Table~\ref{tab:yieldsLLL} and illustrated
in Fig.~\ref{fig:yieldsLLL}, which shows corresponding contributions
in separate channels as well as several characteristic distributions.
The trijet mass for the hadronic top-quark candidate is calculated with the same method as
in Section~\ref{sec:resultsSS}.

The systematic uncertainty in the cross section measurement arises from uncertainties in the background yields and
in the estimate of the signal selection efficiency.
For the signal event selection, the dominant sources of systematic uncertainty are
the modelling of the lepton selection and
the uncertainty in the jet energy scale.
They produce 6\% and 5\% uncertainty in the signal selection efficiency,
respectively, and sum to a total of 10\% systematic uncertainty together with the other sources of uncertainty
described in Section~\ref{sec:signalSyst}.

\begin{table}[h!]
\centering
\topcaption[]{Expected signal, estimated backgrounds, the sum of the two, and observed number of events for
 the trilepton channel. Uncertainties include both the statistical and the systematic components.
 The systematic uncertainty in the signal contribution does not include
 the theoretical uncertainty in the signal production cross section.
}
\label{tab:yieldsLLL}
\begin{tabular}{lr@{$\,\pm\,$}l}
\hline
			          & \multicolumn{2}{c}{Yield}   \\ \hline
 \ttz\ (expected)  & 7.8 & 0.9		\\
\hline
 Irreducible            & 0.8 & 0.4     \\
 \ttw            & 0.2 & 0.1     \\
 Non-top-quark 		& 2.3 & 1.2		\\
 Misidentified lepton   & 1.1 & 0.8		\\
\hline
 Total background	& 4.4 & 1.6		\\
 Total expected		& 12.2 & 1.8		\\
\hline
{Observed}	      & \multicolumn{2}{c}{12}  \\
\hline
\end{tabular}

\end{table}

Based on the observed number of events, the background estimates, and the signal acceptance,
of $0.0021\pm0.0001\stat\pm0.0002\syst$,
the inclusive \ttz\ production cross
section in the trilepton analysis is measured as
\begin{equation*}
\sigma_{\ttz, 3\ell} = 190 ^{+100}_{-80}\stat \pm 40\syst \unit{fb},
\end{equation*}
including statistical and systematic uncertainties,
compared to the SM expectation of  $197^{+22}_{-25}\unit{fb}$.
The significance of the result over the background-only hypothesis is equivalent to 2.3 standard deviations,
compared to the expected value of 2.4.
This result is combined with the four-lepton analysis and the same-sign dilepton analysis, as described in Section~\ref{sec:combination}.

\section{Four-lepton analysis} \label{sec:fourlepton}
The aim of the four-lepton analysis is to select events originating from the process:
\begin{equation*}
\Pp\Pp\to \ttz \to  (\cPqt\to \cPqb\ell\nu)(\cPqt\to \cPqb\ell\nu)(\cPZ \to \ell\overline{\ell}).
\end{equation*}
These events are characterized by a pair of same-flavour, opposite-sign leptons~(\Pe~and \Pgm) with an invariant mass
that is close to the nominal Z-boson mass and two additional prompt leptons.

Since the branching fraction of \ttz to four leptons is very low, it is a challenge to maintain high signal efficiency
and at the same time reject as much background as possible. To that end, the events are separated into two categories, one
of which has a significantly higher signal-to-background ratio than the other. The event selection has been optimized
using the signal significance from simulated events and is summarized in the following:

\begin{enumerate}
	\item Events must have a total of four leptons passing the lepton identification criteria described
        in Section~\ref{sec:evsel}. Each electron (muon) is required to have relative isolation smaller than $9\:(10)\%$.
	\item The highest lepton $\pt$ must be greater than 20\GeV. The remaining leptons must have $\pt > 10$\GeV.
	\item Two of the leptons must form an opposite-sign same-flavour pair with the dilepton mass between 76 and 106\GeV.
	\item The remaining two opposite-sign leptons must not form a same-flavour pair with the
        dilepton mass between 76 and 106\GeV.
	\item At least one jet must pass the medium CSV  b-tagging selection.
	\item At least one other jet must pass the loose CSV b-tagging selection.
\end{enumerate}
The high signal-to-background signal region requires that events pass all of the criteria above. A second signal region requires that they pass the first five conditions and fail the sixth. These two four-lepton channels are exclusive.

\subsection{Background estimation}
\label{sec:L4_bgpred}

The standard model can produce four genuine, prompt leptons through multiboson+jets production where
at least two bosons decay leptonically. Backgrounds to this search include ZZ, WWZ, WZZ, ZZZ, and rarer processes. They can prove irreducible if the multiboson production is accompanied by b-tagged jets arising from the underlying event or
initial-state radiation (\emph{irreducible} background).

The contribution from irreducible background processes is estimated using MC simulations. The process with the largest contribution in the four-lepton signal regions comes from the ZZ process.
The main concern with taking this background estimate solely from a simulation is how well the rate at which bottom quarks are produced is modelled.
Since these bottom quarks mainly originate from initial-state radiation, this rate is estimated in a data sample
of leptonically-decaying Z bosons with two additional jets.
For events in this sample the probability to pass the two b-tagging criteria is found to be about 4\%.
Rescaling by this number the events in the appropriate ZZ enhanced region measured in data, the background estimate
is found to agree very well with the estimate from simulations.
Therefore, the latter estimate is used in the analysis.

Another source of background arises when electrons and muons are incorrectly identified as prompt and
isolated (\emph{misidentified lepton} background). These can either result from misreconstruction of
hadrons or from non-prompt or non-isolated leptons passing the selection criteria.
Isolated tracks are used as a proxy for misidentified leptons and to calculate a ``track-to-lepton'' ratio, which
depends on the heavy-flavour content and jet activity. The track-to-lepton ratio is
determined by measuring the number of prompt,
isolated tracks and the number of prompt, isolated leptons after the contribution to the leptons
from electroweak processes has been subtracted. It is calculated in two control
regions in data: a region with leptonic decays of Z bosons and a region with semi-leptonic decays
of \ttbar pairs. The two regions cover the extremes of how much
heavy-flavour content is expected in different event samples.
The ratio is then interpolated between these two regions using a linear mixing of the two control samples
and parameterized as a function of the variable $R_{\textrm{n-p/p}}$, which is the ratio of non-isolated,
non-prompt tracks to non-isolated, prompt tracks in the sample.
A track is defined as prompt when its transverse impact parameter
is less than 200\micron, and non-prompt otherwise.
The variable $R_{\textrm{n-p/p}}$ is used in the parameterization of the track-to-lepton ratio since
it quantifies the amount of heavy-flavour content in the events of a given sample.
The validity of the parameterization is checked in a
third control region that requires one dilepton pair consistent with the Z boson and at least one b-tagged jet: for this sample,
whose heavy-flavour content is expected to be in between those of the two previous control regions,
$R_{\textrm{n-p/p}}$ is calculated, and the predicted and observed track-to-lepton ratios are compared and
found in agreement.
Finally, two sideband regions with one
dilepton pair consistent with the Z boson and a third lepton, and which also satisfy the two b-tagging
categorizations are defined.
By calculating $R_{\textrm{n-p/p}}$ and using the track-to-lepton parameterization, the probability for isolated,
prompt tracks to be misidentified as electrons (muons) is found equal to
$7.4 \pm 2.2\%$ ($1.6\pm 0.5\%$) in these two samples.
To determine the number of background events in the signal regions, the yields in the
sideband regions are then multiplied by the track-to-lepton ratios and the relevant combinatoric
factors depending on the number of isolated tracks present in the events. A background yield
of $0.1 \pm 0.1$ ($0.5 \pm 0.2$) in the 2 b-jet (1 b-jet) signal region is calculated in this way.

\subsection{Four-lepton results}
\label{sec:tL4_results}

\begin{figure}[htb]
  \centering
  \includegraphics[width=0.45\textwidth]{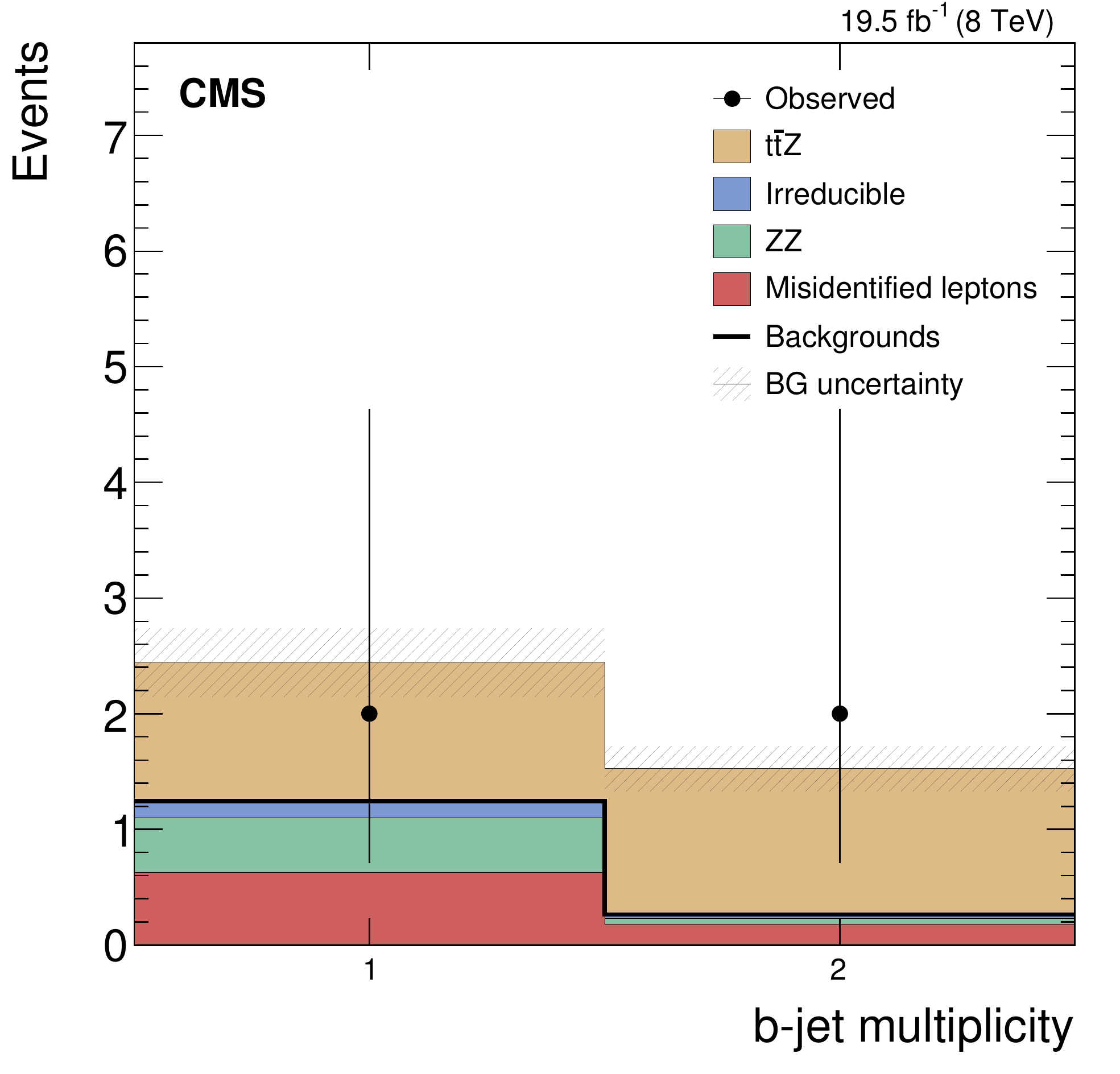}
  \includegraphics[width=0.45\textwidth]{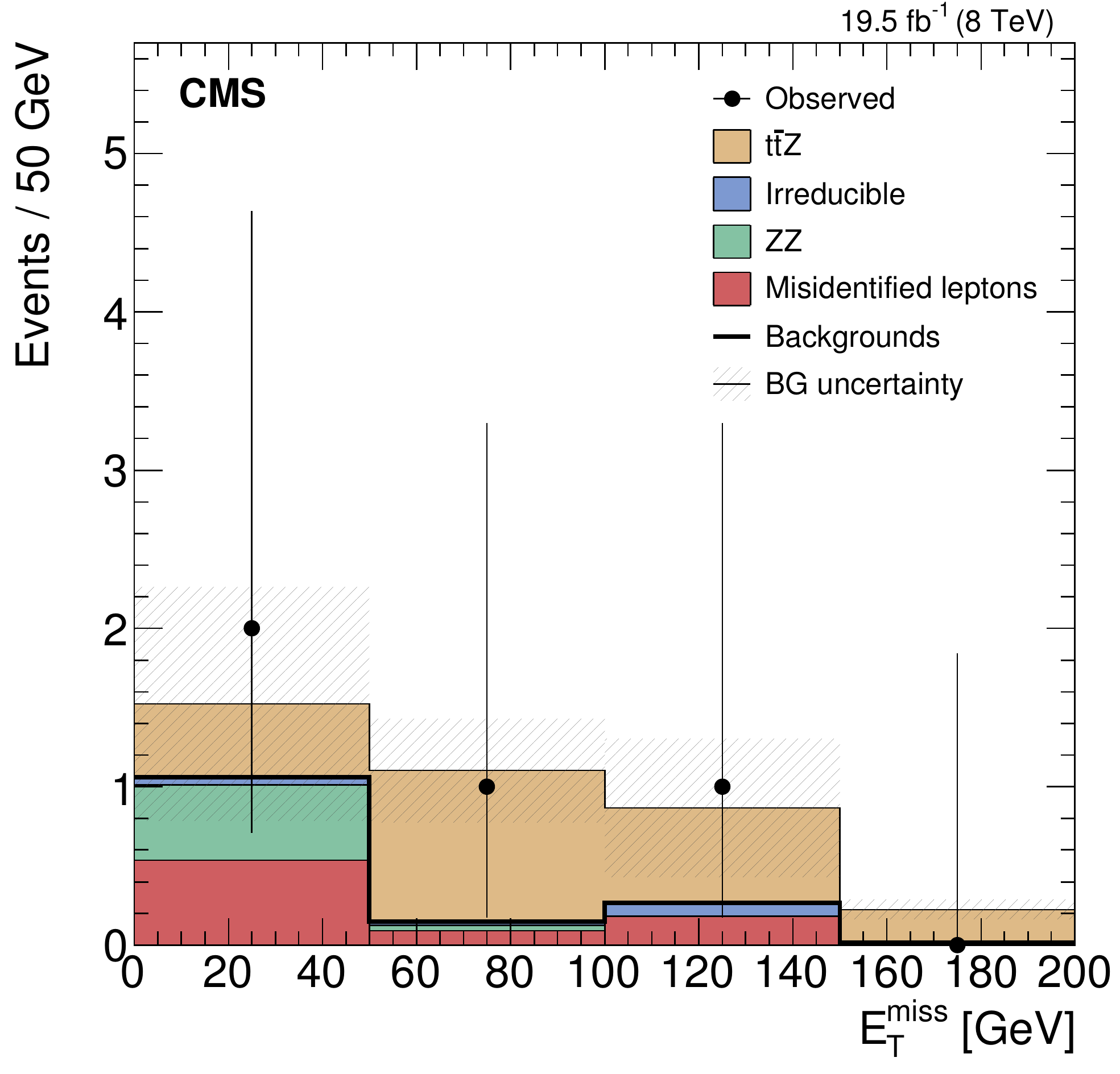}
  \caption{\label{fig:fourLeptonResults} Event yields in data after final four-lepton selection requirements,
    compared to the background estimates  and signal expectations.
    Contributions in the 1~b-tagged jet and 2~b-tagged jets signal regions (\cmsLeft) and inclusive \MET distribution (\cmsRight).
    The ZZ component of the background is shown separately from the rest of the irreducible processes.
    The combination of statistical and systematic uncertainties is denoted by the shaded area.}

\end{figure}

Applying the full event selection, the event yields shown in Table~\ref{tab:fourLeptonResults} are obtained. A total of 4 events are observed, compared to a background expectation of  $1.4 \pm 0.3$ events, where the uncertainty in the background prediction contains both the contributions from the limited number of simulated events and from the uncertainties related to the rescaling procedure based on control samples in data.
The results are shown in Fig.~\ref{fig:fourLeptonResults} (\cmsLeft). A comparison of the \MET distributions for the background, signal, and observed data, combining the two signal regions, is shown in Fig.~\ref{fig:fourLeptonResults} (\cmsRight).

\begin{table}[htb]
\centering
\topcaption{\label{tab:fourLeptonResults} Expected signal, estimated backgrounds, the sum of the two, and observed number of events for
 the four-lepton channel. Uncertainties include both the statistical and the systematic components.
 The systematic uncertainty in the signal contribution does not include
 the theoretical uncertainty in the signal production cross section. The ZZ component of the background is shown separately
 from the rest of the irreducible processes.}
\begin{tabular}{lcc}
\hline
    & 2 b jets required & 1 b jet required \\
\hline
 \ttz (expected) & 1.3 $\pm$ 0.2 & 1.3 $\pm$ 0.2\\
\hline
 Misidentified lepton & 0.1 $\pm$ 0.1 & 0.5 $\pm$ 0.2\\
 ZZ & 0.05 $\pm$ 0.01 & 0.47 $\pm$ 0.02\\
 Irreducible & 0.04 $\pm$ 0.03 & 0.14 $\pm$ 0.04 \\
\hline
 Total background & 0.2 $\pm$ 0.1 & 1.1 $\pm$ 0.2 \\
 Total expected & 1.5 $\pm$ 0.2 & 2.4 $\pm$ 0.3 \\
\hline
 {Observed} & {2} & {2}\\
\hline
\end{tabular}

\end{table}

The systematic uncertainties in the selection efficiencies for signal and irreducible background are derived
in the same way as for the dilepton and trilepton channels and are described in Section~\ref{sec:signalSyst}.
For the four-lepton analysis, the dominant source of uncertainty in the signal acceptance is the
8\% uncertainty in the modelling of the lepton selection. Together with the other systematic uncertainties, it sums
to a total uncertainty of 11\% in the signal selection efficiency.

By performing a simultaneous fit to the two exclusive four-lepton signal regions,
the following cross section is extracted:
\begin{equation*}
\sigma_{\ttz, 4\ell} = 230 ^{+180}_{-130}\stat ^{+60}_{-30}\syst\unit{fb}.
\end{equation*}
The significance is equal to 2.2 standard deviations (2.0 standard deviations expected).

\section{Systematic uncertainties in signal selection efficiency}
\label{sec:signalSyst}

Along with the corresponding techniques for the background estimation, the uncertainties in the estimates of the backgrounds
affecting the three channels have been presented respectively in Sections~\ref{sec:ssdl},
\ref{sec:trilepton}, and \ref{sec:fourlepton}. Here are illustrated the uncertainties in the selection
efficiency of signal events.

Except for the component due to trigger, simulation is used to estimate the selection efficiency for signal.
Control samples in data are used to correct the selection acceptance estimated in simulation
and to assess the corresponding uncertainty. A similar approach is employed for all three analysis channels and therefore
a common list of systematic uncertainties in signal acceptance can be summarized as in
Table~\ref{tab:signalSysUnc}. The total uncertainty in the computed acceptance varies between 8 and 11\% depending on the channel.

\begin{table}[h]
	\centering
		\topcaption{Systematic uncertainties in the signal selection efficiency for
                  the three considered channels: \ttw in dilepton (2$\ell$) final state; \ttz in
                  trilepton  (3$\ell$) and four-lepton (4$\ell$) final states.}
		\label{tab:signalSysUnc}
		\begin{tabular}{ l c c c }  \hline
		                                                   &   \multicolumn{3}{c}{Channels}      \\
		                                                   &\multicolumn{1}{c}{2$\ell$}&\multicolumn{1}{c}{3$\ell$}&\multicolumn{1}{c}{4$\ell$}\\ \hline
                Source of uncertainty                              &   \multicolumn{3}{c}{Uncertainty (\%)}      \\ \hline
		Modelling of trigger eff.                    &   3   &   1   &   1  \\
		Modelling of lepton sel. (ID/isolation)   &   4   &   6   &   8  \\
		Jet energy scale and resolution                    &   4   &   5   &   4  \\
		Identification of b jets                           &   2   &   3   &   3  \\
		Pileup modelling                                   &   1   &   1   &   1  \\
		Choice of parton distribution functions            &  1.5  &  1.5  &  1.5 \\
		Signal model                                       &   5   &   5   &   5  \\  \hline
		Total                                              &   8   &   10  &   11 \\  \hline
		\end{tabular}
	
\end{table}

The trigger efficiency is directly measured in data using control samples selected by \HT triggers that are orthogonal to the
dilepton triggers employed by the three analyses to select signal event candidates~\cite{ssDilepSusy8TeV}. Trigger
inefficiencies are then applied to all acceptances calculated from simulation, for both signal and the background
processes derived from simulation.

The offline lepton selection efficiencies in data and simulation are measured using \cPZ-boson events to derive
simulation-to-data correction factors.
The correction factors applied to simulation are about 0.94 (0.98) for \pt $>$ 20\GeV for electrons (muons).
The uncertainty in the per-lepton selection efficiency is about 1.5\% (0.3\%)
for electrons (muons) with \pt $> 20$\GeV. An additional systematic uncertainty is assigned to account
for potential mismodelling of the lepton
isolation efficiency due to the larger hadronic activity in signal events than in \cPZ-boson events. This uncertainty is in the 2--3\% range.
These per-lepton uncertainties are propagated to calculate the uncertainties in the selection
efficiency of signal events, which are found to be in the 4--8\% range depending on the leptonic final state.

Another source of systematic uncertainty is associated with the
jet energy scale correction. This systematic uncertainty varies between
5\% and 2\% in the \pt range 40--100\GeV for jets with $\abs{\eta}<2.4$~\cite{JES}.
It is evaluated on a single-jet basis, and its effect
is propagated to \HT, the number of jets, and the number of b-tagged jets.
In addition, there is a contribution to the total uncertainty arising from
limited knowledge of the resolution of the jet energy, but this effect is generally
of less importance than the contribution from the jet energy scale.

The b-tagging efficiency for b-quark jets, and the mistagging probabilities for charm-quark jets and for jets originating
from light-flavour quarks or gluons, are estimated from data~\cite{btagging:2012}. The corresponding correction
factors, dependent on jet flavour and kinematic properties, are applied to simulated jets
to account for the differences in the tagging efficiency between simulation and data. The total
uncertainty in the signal acceptance caused by the b-tagging selection is determined by
varying the correction factors up and down by
their uncertainties.

In the simulation of signal events, different pileup conditions have been probed varying  the
cross section for inelastic pp collisions by $\pm$5\%. Comparing the signal selection
efficiency for these different conditions, the uncertainty associated to pileup effects
is found to be approximately 1\%.
The uncertainty in the signal acceptance due to the PDF choice~\cite{Alekhin:2011sk,Botje:2011sn,ct10,nnpdf,MSTW}
is found to be 1.5\%. An uncertainty of the order of 5\% in the signal
acceptance is also assigned to the finite-order calculation employed to generate signal events. This last uncertainty, which
covers also the uncertainty in the effects of initial- and final-state radiation, is estimated varying
from their nominal values the matrix-element/parton-shower matching scale (with the nominal value of 20\GeV), and the
renormalization and factorization scales (with the nominal value equal to $Q^2$ in the event).
For the up and down variations of the matching scale, thresholds of 40 and 10\GeV are used,
respectively. Renormalization and factorization scales are varied between 4Q$^2$ and Q$^2$/4.
The signal model uncertainty also includes the difference in acceptance between signal events simulated
with \MADGRAPH{}5 and a\MCATNLO~\cite{amcatnlo} generators.

\section{Results} \label{sec:combination}
To extract the cross sections for the \ttw and \ttz~processes,
the nine different channels are combined to maximize their sensitivity.
Cross section central values and corresponding uncertainties are evaluated from a scan of the profile
likelihood ratio.
The adopted statistical procedure is the same that was used for the observation of the Higgs boson candidate in CMS, and
is described in detail in Ref.~\cite{HiggsObservationCMS}.

{\tolerance=900
The results of the measurements are summarized in Table~\ref{tab:combination}. Two one-dimensional fits are performed to measure \ttw
and \ttz separately using the channels most sensitive to each process.
Using only the same-sign dilepton channels, the extracted \ttw cross section is measured to be
$170 ^{+90}_{-80}\stat \pm 70\syst$\unit{fb}, corresponding to a
significance of $1.6$ standard deviations over the background-only hypothesis.
The three and four lepton channels are combined to extract a \ttz cross section of
$200 ^{+80}_{-70}\stat ^{+40}_{-30}\syst$\unit{fb}, with a significance of 3.1 standard deviations.
\par}

\begin{table*}[htb]
\renewcommand{\arraystretch}{1.35}
\centering
\topcaption{\label{tab:combination} Results of the extraction of cross sections, from single and combined channels.
                                 The significance is expressed in terms of standard deviations.}
\begin{tabular}{cccc}
\hline
Channels used & Process &  Cross section & Significance  \\
\hline

2$\ell$ &  \ttw & $170 ^{+90}_{-80}\stat \pm 70\syst$\unit{fb} & 1.6 \\
3$\ell$+4$\ell$ &  \ttz & $200 ^{+80}_{-70}\stat ^{+40}_{-30}\syst$\unit{fb} & 3.1 \\
2$\ell$+3$\ell$+4$\ell$ & \ttw $+$ \ttz & $380 ^{+100}_{-90}\stat ^{+80}_{-70}\syst$\unit{fb} & 3.7 \\
\hline
\end{tabular}

\end{table*}

When calculating the one-dimensional fit of one process, the cross section of the other process
is constrained to have the theoretical SM value with a systematic uncertainty of 50\%.

\begin{figure}[htb]
\centering
\includegraphics[width=\cmsFigWidthLarge]{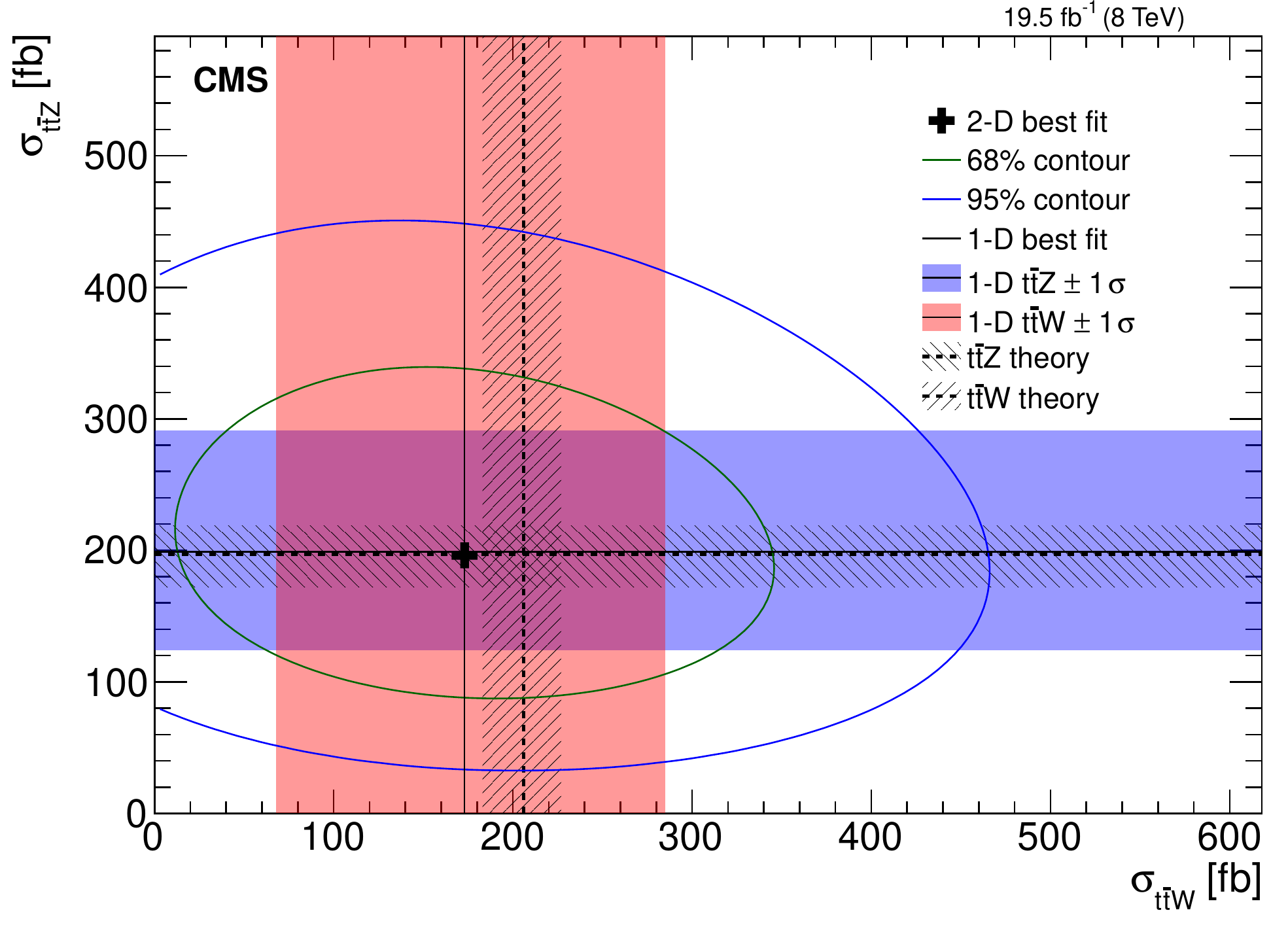}
\caption{\label{fig:combination_2d} The result of the two-dimensional best fit for \ttw and \ttz{} cross sections
(cross symbol)
is shown along with its 68 and 95\% confidence level contours. The result of this fit is superimposed with the
separate \ttw and \ttz{} cross section measurements, and the corresponding 1 standard deviation (1$\sigma$) bands, obtained
from the dilepton, and the trilepton/four-lepton channels, respectively. The figure also shows the
predictions from theory and the corresponding uncertainties.}
\end{figure}

As visible from Fig.~\ref{fig:yieldSS} and Table~\ref{tab:yieldSS}, less than 10\% of the events selected by the
same-sign dilepton channels are expected to stem from \ttz production.
The extracted \ttw cross section varies by approximately 10\%
when the used \ttz cross section is altered to as much as 0.5--1.5 times its nominal theoretical value.
For an equivalent modification of the \ttw production rate, the variation of the extracted \ttz cross section
is less than 2\%.
The dependence of the measured cross section on the assumed cross section of the other \ttv process
is solved by performing a simultaneous fit of the cross sections of the two processes using all dilepton,
trilepton, and four-lepton channels at the same time.

\begin{table}[htb]
\centering
\topcaption{\label{tab:fit2d} Results for the two dimensional fit of the \ttw and \ttz cross sections.}
\begin{tabular}{ccc}
\hline

Channels used & \ttw  cross section &  \ttz{} cross section \\
\hline

2$\ell$+3$\ell$+4$\ell$ & $170 ^{+110}_{-100}\,\text{(total)}$\unit{fb} & $200 \pm 90\,\text{(total)}$\unit{fb}  \\

\hline

\end{tabular}

\end{table}

The result of the fit is shown visually in Fig.~\ref{fig:combination_2d} and the cross sections
are summarized numerically in Table~\ref{tab:fit2d}. The cross sections extracted from this two-dimensional
fit are identical to those obtained from the two one-dimensional fits.

{\tolerance=1200
Finally, a one-dimensional fit of all channels is performed to extract a combined cross section
$\sigma_{\ttv} = 380 ^{+100}_{-90}\stat ^{+80}_{-70}\syst$\unit{fb}
with a significance of 3.7 standard deviations.
\par}

\section{Summary}

A measurement with the CMS detector of the cross section of top quark-antiquark pair production in association
with a W or Z boson at $\sqrt{s} = 8$\TeV has been presented.
Results from three independent channels, and their combination, have been reported.
In the same-sign dilepton channel, the \ttw cross section has been measured to be $\sigma_{\ttw} = 170  ^{+90}_{-80}\stat \pm 70\syst\unit{fb}$, corresponding to a significance of 1.6 standard deviations
over the background-only hypothesis.
In the trilepton and four-lepton channels the \ttz signal has been established with a significance of 2.3 and 2.2 standard
deviations, respectively. From the combination of these two channels, a significance of 3.1 standard deviations
has been obtained and the cross section has been measured to be
$\sigma_{\ttz} = 200  ^{+80}_{-70}\stat ^{+40}_{-30}\syst\unit{fb}$.

Combining the total of nine sub-channels from the three lepton decay modes, a \ttv cross section (V equal W or Z) of
$ \sigma_{\ttv} = 380  ^{+100}_{-90}\stat ^{+80}_{-70}\syst\unit{fb}$
has been obtained, corresponding to a combined significance of 3.7 standard deviations.
The measured values are compatible within their uncertainties with standard model predictions.

\section*{Acknowledgements}
{\tolerance=1200
We congratulate our colleagues in the CERN accelerator departments for the excellent performance of the LHC and thank the technical and administrative staffs at CERN and at other CMS institutes for their contributions to the success of the CMS effort. In addition, we gratefully acknowledge the computing centres and personnel of the Worldwide LHC Computing Grid for delivering so effectively the computing infrastructure essential to our analyses. Finally, we acknowledge the enduring support for the construction and operation of the LHC and the CMS detector provided by the following funding agencies: BMWFW and FWF (Austria); FNRS and FWO (Belgium); CNPq, CAPES, FAPERJ, and FAPESP (Brazil); MES (Bulgaria); CERN; CAS, MoST, and NSFC (China); COLCIENCIAS (Colombia); MSES and CSF (Croatia); RPF (Cyprus); MoER, ERC IUT and ERDF (Estonia); Academy of Finland, MEC, and HIP (Finland); CEA and CNRS/IN2P3 (France); BMBF, DFG, and HGF (Germany); GSRT (Greece); OTKA and NIH (Hungary); DAE and DST (India); IPM (Iran); SFI (Ireland); INFN (Italy); NRF and WCU (Republic of Korea); LAS (Lithuania); MOE and UM (Malaysia); CINVESTAV, CONACYT, SEP, and UASLP-FAI (Mexico); MBIE (New Zealand); PAEC (Pakistan); MSHE and NSC (Poland); FCT (Portugal); JINR (Dubna); MON, RosAtom, RAS and RFBR (Russia); MESTD (Serbia); SEIDI and CPAN (Spain); Swiss Funding Agencies (Switzerland); MST (Taipei); ThEPCenter, IPST, STAR and NSTDA (Thailand); TUBITAK and TAEK (Turkey); NASU and SFFR (Ukraine); STFC (United Kingdom); DOE and NSF (USA).

Individuals have received support from the Marie-Curie programme and the European Research Council and EPLANET (European Union); the Leventis Foundation; the A. P. Sloan Foundation; the Alexander von Humboldt Foundation; the Belgian Federal Science Policy Office; the Fonds pour la Formation \`a la Recherche dans l'Industrie et dans l'Agriculture (FRIA-Belgium); the Agentschap voor Innovatie door Wetenschap en Technologie (IWT-Belgium); the Ministry of Education, Youth and Sports (MEYS) of the Czech Republic; the Council of Science and Industrial Research, India; the HOMING PLUS programme of Foundation for Polish Science, cofinanced from European Union, Regional Development Fund; the Compagnia di San Paolo (Torino); the Thalis and Aristeia programmes cofinanced by EU-ESF and the Greek NSRF; and the National Priorities Research Program by Qatar National Research Fund.
\par}

\bibliography{auto_generated}   % will be created by the tdr script.

\providecommand{\href}[2]{#2}\begingroup\raggedright\begin{thebibliography}{10}%
\makeatletter
\providecommand{\hrefCMSnoop }[0]{\@secondoftwo}%
\makeatother
\providecommand{\doi}{\texttt{doi:}\begingroup \urlstyle{tt}\Url}

\bibitem{top_cdf}
\hrefCMSnoop {} {{ CDF} Collaboration, ``Observation of Top Quark Production in
  $\overline{\mathit{p}}\mathit{p}$ Collisions with the Collider Detector at
  Fermilab'',} \textit{ Phys. Rev. Lett.} \textbf{ 74} (1995) 2626,
  \href{http://dx.doi.org/10.1103/PhysRevLett.74.2626}{\doi{10.1103/PhysRevLett.74.2626}},
  \href{http://www.arXiv.org/abs/hep-ex/9503002}{\texttt{
  arXiv:hep-ex/9503002}}.

\bibitem{top_d0_discovery}
\hrefCMSnoop {} {{ D0} Collaboration, ``Observation of the Top Quark'',}
  \textit{ Phys. Rev. Lett.} \textbf{ 74} (1995) 2632,
  \href{http://dx.doi.org/10.1103/PhysRevLett.74.2632}{\doi{10.1103/PhysRevLett.74.2632}}.

\bibitem{sm_Glashow}
\hrefCMSnoop {} {S.~L. Glashow, ``Partial-symmetries of weak interactions'',}
  \textit{ Nucl. Phys.} \textbf{ 22} (1961) 579,
\href{http://dx.doi.org/10.1016/0029-5582(61)90469-2}{\doi{10.1016/0029-5582(61)90469-2}}.
%%CITATION = NUPHA,22,579;%%.

\bibitem{sm_Weinberg}
\hrefCMSnoop {} {S.~Weinberg, ``{A Model of Leptons}'',} \textit{ Phys. Rev.
  Lett.} \textbf{ 19} (1967) 1264,
\href{http://dx.doi.org/10.1103/PhysRevLett.19.1264}{\doi{10.1103/PhysRevLett.19.1264}}.
%%CITATION = PRLTA,19,1264;%%.

\bibitem{sm_Salam}
\hrefCMSnoop {} {A.~Salam, ``Weak and electromagnetic interactions'',} in
  \textit{ Elementary particle physics: relativistic groups and analyticity},
  N.~Svartholm, ed., p.~367.
\newblock Almqvist \& Wiskell, 1968.
\newblock Proceedings of the eighth Nobel symposium.

\bibitem{Barnett:1993ea}
\hrefCMSnoop {} {R.~M. Barnett, J.~F. Gunion, and H.~E. Haber, ``{Discovering
  supersymmetry with like-sign dileptons}'',} \textit{ Phys. Lett. B} \textbf{
  315} (1993) 349,
  \href{http://dx.doi.org/10.1016/0370-2693(93)91623-U}{\doi{10.1016/0370-2693(93)91623-U}},
\href{http://www.arXiv.org/abs/hep-ph/9306204}{\texttt{ arXiv:hep-ph/9306204}}.
%%CITATION = HEP-PH/9306204;%%.

\bibitem{Guchait:1994zk}
\hrefCMSnoop {} {M.~Guchait and D.~P. Roy, ``{Like-sign dilepton signature for
  gluino production at CERN LHC including top quark and Higgs boson
  effects}'',} \textit{ Phys. Rev. D} \textbf{ 52} (1995) 133,
  \href{http://dx.doi.org/10.1103/PhysRevD.52.133}{\doi{10.1103/PhysRevD.52.133}},
\href{http://www.arXiv.org/abs/hep-ph/9412329}{\texttt{ arXiv:hep-ph/9412329}}.
%%CITATION = HEP-PH/9412329;%%.

\bibitem{Baer:1995va}
\hrefCMSnoop {} {H.~Baer, C.-h. Chen, F.~Paige, and X.~Tata, ``{Signals for
  minimal supergravity at the CERN large hadron collider. II: Multi-lepton
  channels}'',} \textit{ Phys. Rev. D} \textbf{ 53} (1996) 6241,
  \href{http://dx.doi.org/10.1103/PhysRevD.53.6241}{\doi{10.1103/PhysRevD.53.6241}},
\href{http://www.arXiv.org/abs/hep-ph/9512383}{\texttt{ arXiv:hep-ph/9512383}}.
%%CITATION = HEP-PH/9512383;%%.

\bibitem{ssDilepSusy8TeV}
\hrefCMSnoop {} {{ CMS} Collaboration, ``{Search for new physics in events with
  same-sign dileptons and jets in pp collisions at $\sqrt{s}$ = 8 TeV}'',}
  \textit{ JHEP} \textbf{ 01} (2014) 163,
  \href{http://dx.doi.org/10.1007/JHEP01(2014)163}{\doi{10.1007/JHEP01(2014)163}},
\href{http://www.arXiv.org/abs/1311.6736}{\texttt{ arXiv:1311.6736}}.
%%CITATION = ARXIV:1311.6736;%%.

\bibitem{multileptonSearch7TeV}
\hrefCMSnoop {} {{ CMS} Collaboration, ``{Search for anomalous production of
  multilepton events in pp collisions at $\sqrt{s}=7$ TeV}'',} \textit{ JHEP}
  \textbf{ 06} (2012) 169,
  \href{http://dx.doi.org/10.1007/JHEP06(2012)169}{\doi{10.1007/JHEP06(2012)169}},
\href{http://www.arXiv.org/abs/1204.5341}{\texttt{ arXiv:1204.5341}}.
%%CITATION = ARXIV:1204.5341;%%.

\bibitem{ttgCDF}
\hrefCMSnoop {} {{ CDF} Collaboration, ``{Evidence for
  $\textrm{t}\bar{\textrm{t}}\gamma$ Production and Measurement of
  $\sigma_{\textrm{t}\bar{\textrm{t}}\gamma} /
  \sigma_{\textrm{t}\bar{\textrm{t}}}$}'',} \textit{ Phys. Rev. D} \textbf{ 84}
  (2011) 031104,
  \href{http://dx.doi.org/10.1103/PhysRevD.84.031104}{\doi{10.1103/PhysRevD.84.031104}},
\href{http://www.arXiv.org/abs/1106.3970}{\texttt{ arXiv:1106.3970}}.
%%CITATION = ARXIV:1106.3970;%%.

\bibitem{ttVat7TeV}
\hrefCMSnoop {} {{ CMS} Collaboration, ``{Measurement of associated production
  of vector bosons and top quark-antiquark pairs at $\sqrt{s}$ = 7 TeV}'',}
  \textit{ Phys. Rev. Lett.} \textbf{ 110} (2013) 172002,
  \href{http://dx.doi.org/10.1103/PhysRevLett.110.172002}{\doi{10.1103/PhysRevLett.110.172002}},
\href{http://www.arXiv.org/abs/1303.3239}{\texttt{ arXiv:1303.3239}}.
%%CITATION = ARXIV:1303.3239;%%.

\bibitem{ct10}
H.-L. Lai\hrefCMSnoop {} { {et~al.}, ``New parton distributions for collider
  physics'',} \textit{ Phys. Rev. D} \textbf{ 82} (2010) 074024,
  \href{http://dx.doi.org/10.1103/PhysRevD.82.074024}{\doi{10.1103/PhysRevD.82.074024}},
\href{http://www.arXiv.org/abs/1007.2241}{\texttt{ arXiv:1007.2241}}.
%%CITATION = ARXIV:1007.2241;%%.

\bibitem{MG5NLO}
J.~Alwall\hrefCMSnoop {} { {et~al.}, ``The automated computation of tree-level
  and next-to-leading order differential cross sections, and their matching to
  parton shower simulations'',} \textit{ JHEP} \textbf{ 07} (2014) 079,
  \href{http://dx.doi.org/10.1007/JHEP07(2014)079}{\doi{10.1007/JHEP07(2014)079}},
\href{http://www.arXiv.org/abs/1405.0301}{\texttt{ arXiv:1405.0301}}.
%%CITATION = ARXIV:1405.0301;%%.

\bibitem{amcatnlo}
\hrefCMSnoop {} {S.~Frixione and B.~R. Webber, ``{Matching NLO QCD computations
  and parton shower simulations}'',} \textit{ JHEP} \textbf{ 06} (2002) 029,
  \href{http://dx.doi.org/10.1088/1126-6708/2002/06/029}{\doi{10.1088/1126-6708/2002/06/029}},
\href{http://www.arXiv.org/abs/hep-ph/0204244}{\texttt{ arXiv:hep-ph/0204244}}.
%%CITATION = HEP-PH/0204244;%%.

\bibitem{ttWxsecNLO}
\hrefCMSnoop {} {J.~M. Campbell and R.~K. Ellis,
  ``{$\textrm{t}\bar{\textrm{t}}\textrm{W}^{\pm}$ production and decay at
  NLO}'',} \textit{ JHEP} \textbf{ 07} (2012) 052,
  \href{http://dx.doi.org/10.1007/JHEP07(2012)052}{\doi{10.1007/JHEP07(2012)052}},
\href{http://www.arXiv.org/abs/1204.5678}{\texttt{ arXiv:1204.5678}}.
%%CITATION = ARXIV:1204.5678;%%.

\bibitem{ttZxsecNLO}
\hrefCMSnoop {} {M.~V. Garzelli, A.~Kardos, C.~G. Papadopoulos, and
  Z.~Trocsanyi, ``{$\textrm{t}\bar{\textrm{t}}\textrm{W}^{\pm}$ and
  $\textrm{t}\bar{\textrm{t}}\textrm{Z}$ hadroproduction at NLO accuracy in QCD
  with Parton Shower and Hadronization effects}'',} \textit{ JHEP} \textbf{ 11}
  (2012) 056,
  \href{http://dx.doi.org/10.1007/JHEP11(2012)056}{\doi{10.1007/JHEP11(2012)056}},
\href{http://www.arXiv.org/abs/1208.2665}{\texttt{ arXiv:1208.2665}}.
%%CITATION = ARXIV:1208.2665;%%.

\bibitem{cmsJINST:2008}
\hrefCMSnoop {} {{ CMS} Collaboration, ``The {CMS} experiment at the {CERN}
  {LHC}'',} \textit{ JINST} \textbf{ 3} (2008) S08004,
  \href{http://dx.doi.org/10.1088/1748-0221/3/08/S08004}{\doi{10.1088/1748-0221/3/08/S08004}}.

\bibitem{CMS-PAS-PFT-09-001}
\href {http://cdsweb.cern.ch/record/1194487} {{ CMS} Collaboration,
  ``Particle--Flow Event Reconstruction in {CMS} and Performance for Jets,
  Taus, and {\MET}'',} CMS Physics Analysis Summary CMS-PAS-PFT-09-001, 2009.

\bibitem{CMS-PAS-PFT-10-001}
\href {http://cdsweb.cern.ch/record/1247373} {{ CMS} Collaboration,
  ``Commissioning of the Particle-flow Event Reconstruction with the first
  {LHC} collisions recorded in the {CMS} detector'',} CMS Physics Analysis
  Summary CMS-PAS-PFT-10-001, 2010.

\bibitem{antiKT:2008}
\hrefCMSnoop {} {M.~Cacciari, G.~P. Salam, and G.~Soyez, ``The anti-$k_t$ jet
  clustering algorithm'',} \textit{ JHEP} \textbf{ 04} (2008) 063,
  \href{http://dx.doi.org/10.1088/1126-6708/2008/04/063}{\doi{10.1088/1126-6708/2008/04/063}},
  \href{http://www.arXiv.org/abs/0802.1189}{\texttt{ arXiv:0802.1189}}.

\bibitem{JES}
\hrefCMSnoop {} {{ CMS} Collaboration, ``Determination of jet energy
  calibration and transverse momentum resolution in {CMS}'',} \textit{ JINST}
  \textbf{ 6} (2011) P11002,
  \href{http://dx.doi.org/10.1088/1748-0221/6/11/P11002}{\doi{10.1088/1748-0221/6/11/P11002}},
  \href{http://www.arXiv.org/abs/1107.4277}{\texttt{ arXiv:1107.4277}}.

\bibitem{Cacciari:2008gn}
\hrefCMSnoop {} {M.~Cacciari, G.~P. Salam, and G.~Soyez, ``The catchment area
  of jets'',} \textit{ JHEP} \textbf{ 04} (2008) 005,
  \href{http://dx.doi.org/10.1088/1126-6708/2008/04/005}{\doi{10.1088/1126-6708/2008/04/005}},
\href{http://www.arXiv.org/abs/0802.1188}{\texttt{ arXiv:0802.1188}}.
%%CITATION = 0802.1188;%%.

\bibitem{generalJetId}
\hrefCMSnoop {} {{ CMS} Collaboration, ``{Measurements of differential jet
  cross sections in proton-proton collisions at $\sqrt{s}=7$ TeV with the CMS
  detector}'',} \textit{ Phys. Rev. D} \textbf{ 87} (2013) 112002,
  \href{http://dx.doi.org/10.1103/PhysRevD.87.112002}{\doi{10.1103/PhysRevD.87.112002}},
\href{http://www.arXiv.org/abs/1212.6660}{\texttt{ arXiv:1212.6660}}.
%%CITATION = ARXIV:1212.6660;%%.

\bibitem{jetIDPU}
\href {http://cdsweb.cern.ch/record/1581583} {{ CMS} Collaboration, ``{Pileup
  Jet Identification}'',} CMS Physics Analysis Summary CMS-PAS-JME-13-005,
  2013.

\bibitem{btagging:8TevPAS}
\href {http://cds.cern.ch/record/1581306} {{ CMS} Collaboration, ``{Performance
  of b tagging at $\sqrt{s}$=8 TeV in multijet, ttbar and boosted topology
  events}'',} CMS Physics Analysis Summary CMS-PAS-BTV-13-001, 2013.

\bibitem{MUOPAS}
\hrefCMSnoop {} {{ CMS} Collaboration, ``{Performance of CMS muon
  reconstruction in pp collision events at $\sqrt{s}=7$ TeV}'',} \textit{
  JINST} \textbf{ 7} (2012) P10002,
  \href{http://dx.doi.org/10.1088/1748-0221/7/10/P10002}{\doi{10.1088/1748-0221/7/10/P10002}},
\href{http://www.arXiv.org/abs/1206.4071}{\texttt{ arXiv:1206.4071}}.
%%CITATION = ARXIV:1206.4071;%%.

\bibitem{EGMPAS}
\href {http://cdsweb.cern.ch/record/1299116} {{ CMS} Collaboration, ``Electron
  Reconstruction and Identification at $\sqrt{s} = 7$ {TeV}'',} CMS Physics
  Analysis Summary CMS-PAS-EGM-10-004, 2010.

\bibitem{HiggsObservationCMS}
\hrefCMSnoop {} {{ CMS} Collaboration, ``{Observation of a new boson with mass
  near 125 GeV in pp collisions at $\sqrt{s}$ = 7 and 8 TeV}'',} \textit{ JHEP}
  \textbf{ 06} (2013) 081,
  \href{http://dx.doi.org/10.1007/JHEP06(2013)081}{\doi{10.1007/JHEP06(2013)081}},
\href{http://www.arXiv.org/abs/1303.4571}{\texttt{ arXiv:1303.4571}}.
%%CITATION = ARXIV:1303.4571;%%.

\bibitem{MADGRAPH5}
J.~Alwall\hrefCMSnoop {} { {et~al.}, ``{\MADGRAPH 5: going beyond}'',} \textit{
  JHEP} \textbf{ 06} (2011) 128,
  \href{http://dx.doi.org/10.1007/JHEP06(2011)128}{\doi{10.1007/JHEP06(2011)128}},
\href{http://www.arXiv.org/abs/1106.0522}{\texttt{ arXiv:1106.0522}}.
%%CITATION = ARXIV:1106.0522;%%.

\bibitem{PYTHIA}
\hrefCMSnoop {} {T.~Sj{\"o}strand, S.~Mrenna, and P.~Z. Skands, ``{\PYTHIA 6.4
  physics and manual}'',} \textit{ JHEP} \textbf{ 05} (2006) 026,
  \href{http://dx.doi.org/10.1088/1126-6708/2006/05/026}{\doi{10.1088/1126-6708/2006/05/026}},
\href{http://www.arXiv.org/abs/hep-ph/0603175}{\texttt{ arXiv:hep-ph/0603175}}.
%%CITATION = HEP-PH/0603175;%%.

\bibitem{Geant}
\hrefCMSnoop {} {{ GEANT4} Collaboration, ``{\GEANTfour} -- a simulation
  toolkit'',} \textit{ Nucl. Instrum. Meth. A} \textbf{ 506} (2003) 250,
  \href{http://dx.doi.org/10.1016/S0168-9002(03)01368-8}{\doi{10.1016/S0168-9002(03)01368-8}}.

\bibitem{cteq}
P.~M. Nadolsky\hrefCMSnoop {} { {et~al.}, ``{Implications of CTEQ global
  analysis for collider observables}'',} \textit{ Phys. Rev. D} \textbf{ 78}
  (2008) 013004,
  \href{http://dx.doi.org/10.1103/PhysRevD.78.013004}{\doi{10.1103/PhysRevD.78.013004}},
\href{http://www.arXiv.org/abs/0802.0007}{\texttt{ arXiv:0802.0007}}.
%%CITATION = ARXIV:0802.0007;%%.

\bibitem{Beringer:1900zz}
\hrefCMSnoop {} {{ Particle Data Group} Collaboration, ``{Review of Particle
  Physics}'',} \textit{ Phys. Rev. D} \textbf{ 86} (2012) 010001,
\href{http://dx.doi.org/10.1103/PhysRevD.86.010001}{\doi{10.1103/PhysRevD.86.010001}}.
%%CITATION = PHRVA,D86,010001;%%.

\bibitem{DPS}
\hrefCMSnoop {} {{ CMS} Collaboration, ``{Study of double parton scattering
  using W + 2-jet events in proton-proton collisions at $\sqrt{s}$ = 7 TeV}'',}
  \textit{ JHEP} \textbf{ 03} (2014) 032,
  \href{http://dx.doi.org/10.1007/JHEP03(2014)032}{\doi{10.1007/JHEP03(2014)032}},
\href{http://www.arXiv.org/abs/1312.5729}{\texttt{ arXiv:1312.5729}}.
%%CITATION = ARXIV:1312.5729;%%.

\bibitem{CMS-PAS-LUM-13-001}
\href {http://cdsweb.cern.ch/record/1598864} {{ CMS} Collaboration, ``CMS
  Luminosity Based on Pixel Cluster Counting - Summer 2013 Update'',} CMS
  Physics Analysis Summary CMS-PAS-LUM-13-001, 2013.

\bibitem{BCorrelation}
\hrefCMSnoop {} {{ CMS} Collaboration, ``{Measurement of $B\bar{B}$ Angular
  Correlations based on Secondary Vertex Reconstruction at $\sqrt{s}=7$
  TeV}'',} \textit{ JHEP} \textbf{ 03} (2011) 136,
  \href{http://dx.doi.org/10.1007/JHEP03(2011)136}{\doi{10.1007/JHEP03(2011)136}},
\href{http://www.arXiv.org/abs/1102.3194}{\texttt{ arXiv:1102.3194}}.
%%CITATION = ARXIV:1102.3194;%%.

\bibitem{btagging:2012}
\hrefCMSnoop {} {{ CMS} Collaboration, ``{Identification of b-quark jets with
  the CMS experiment}'',} \textit{ JINST} \textbf{ 8} (2013) P04013,
  \href{http://dx.doi.org/10.1088/1748-0221/8/04/P04013}{\doi{10.1088/1748-0221/8/04/P04013}},
\href{http://www.arXiv.org/abs/1211.4462}{\texttt{ arXiv:1211.4462}}.
%%CITATION = ARXIV:1211.4462;%%.

\bibitem{Alekhin:2011sk}
\hrefCMSnoop {} {S.~Alekhin {et~al.}, ``{The PDF4LHC Working Group Interim
  Report}'',} (2011).
\href{http://www.arXiv.org/abs/1101.0536}{\texttt{ arXiv:1101.0536}}.
%%CITATION = ARXIV:1101.0536;%%.

\bibitem{Botje:2011sn}
M.~Botje\hrefCMSnoop {} { {et~al.}, ``{The PDF4LHC Working Group Interim
  Recommendations}'',} (2011).
\href{http://www.arXiv.org/abs/1101.0538}{\texttt{ arXiv:1101.0538}}.
%%CITATION = ARXIV:1101.0538;%%.

\bibitem{nnpdf}
\hrefCMSnoop {} {{ NNPDF} Collaboration, ``Impact of heavy quark masses on
  parton distributions and {LHC} Phenomenology'',} \textit{ Nucl. Phys. B}
  \textbf{ 849} (2011) 296,
  \href{http://dx.doi.org/10.1016/j.nuclphysb.2011.03.021}{\doi{10.1016/j.nuclphysb.2011.03.021}},
\href{http://www.arXiv.org/abs/1101.1300}{\texttt{ arXiv:1101.1300}}.
%%CITATION = ARXIV:1101.1300;%%.

\bibitem{MSTW}
\hrefCMSnoop {} {A.~D. Martin, W.~J. Stirling, R.~S. Thorne, and G.~Watt,
  ``Parton distributions for the {LHC}'',} \textit{ Eur. Phys. J. C} \textbf{
  63} (2009) 189,
  \href{http://dx.doi.org/10.1140/epjc/s10052-009-1072-5}{\doi{10.1140/epjc/s10052-009-1072-5}},
  \href{http://www.arXiv.org/abs/0901.0002}{\texttt{ arXiv:0901.0002}}.

\end{thebibliography}\endgroup

\cleardoublepage \appendix\section{The CMS Collaboration \label{app:collab}}\begin{sloppypar}\hyphenpenalty=5000\widowpenalty=500\clubpenalty=5000\textbf{Yerevan Physics Institute,  Yerevan,  Armenia}\\*[0pt]
V.~Khachatryan, A.M.~Sirunyan, A.~Tumasyan
\vskip\cmsinstskip
\textbf{Institut f\"{u}r Hochenergiephysik der OeAW,  Wien,  Austria}\\*[0pt]
W.~Adam, T.~Bergauer, M.~Dragicevic, J.~Er\"{o}, C.~Fabjan\cmsAuthorMark{1}, M.~Friedl, R.~Fr\"{u}hwirth\cmsAuthorMark{1}, V.M.~Ghete, C.~Hartl, N.~H\"{o}rmann, J.~Hrubec, M.~Jeitler\cmsAuthorMark{1}, W.~Kiesenhofer, V.~Kn\"{u}nz, M.~Krammer\cmsAuthorMark{1}, I.~Kr\"{a}tschmer, D.~Liko, I.~Mikulec, D.~Rabady\cmsAuthorMark{2}, B.~Rahbaran, H.~Rohringer, R.~Sch\"{o}fbeck, J.~Strauss, A.~Taurok, W.~Treberer-Treberspurg, W.~Waltenberger, C.-E.~Wulz\cmsAuthorMark{1}
% ...
\vskip\cmsinstskip
% ...
1:~~Also at Vienna University of Technology, Vienna, Austria\\
2:~~Also at CERN, European Organization for Nuclear Research, Geneva, Switzerland\\
% ...

\end{sloppypar}
\end{document}